\documentclass[fleqn,usenatbib]{mnras}

\usepackage{siunitx}

\usepackage{graphicx}	
\usepackage{amssymb}	
\usepackage{pifont}
\usepackage{makecell}
\usepackage{booktabs}
\usepackage{multirow}
\usepackage{mathtools}
\usepackage{xcolor}
\usepackage{newtxtext,newtxmath}  
\usepackage{float}

\defcitealias{Yu2020}{Y20}

\title[Starburst galaxy outflows and their X-rays]{Outflows from 
  starburst galaxies with various driving mechanisms 
  and their X-ray properties}

\author[Yu, Owen, Pan, Wu \& Ferreras]{
B. P. Brian Yu$^{1,2}$\thanks{E-mail: brian.yu.16@ucl.ac.uk (BPBY), erowen@gapp.nthu.edu.tw (ERO), kuochuan.pan@gapp.nthu.edu.tw (KCP),  
kinwah.wu@ucl.ac.uk (KW)}, 
Ellis R. Owen$^{1,3}$,
Kuo-Chuan Pan$^{1,3}$,
Kinwah Wu$^{2,4}$ and 
\newauthor
\;\! Ignacio Ferreras$^{5,6,7}$
\vspace*{0.25cm} \\
$^{1}$Institute of Astronomy, National Tsing Hua University, Hsinchu, Taiwan (ROC) \\
$^{2}$Mullard Space Science Laboratory, University College London, Dorking, Surrey, RH5 6NT, United Kingdom\\
$^{3}$Center for Informatics and Computation in Astronomy, National Tsing Hua University, Hsinchu, Taiwan (ROC)\\
$^{4}$Research Center for Astronomy, Astrophysics and Astrophotonics, Macquarie University, Sydney, NSW 2019, Australia \\ 
$^{5}$Department of Physics and Astronomy, University College London,
Gower Street, London WC1E 6BT, United Kingdom\\
$^{6}$Instituto de Astrof\'isica de Canarias, C/V\'ia L\'actea, s/n, E38205 La Laguna, Tenerife, Spain\\
$^{7}$Departamento de Astrof{\'i}sica, Universidad de La Laguna (ULL), E-38206 La Laguna, Tenerife, Spain
}

\date{Accepted XXX. Received YYY; in original form ZZZ}

\pubyear{2021}

\begin{document}
\label{firstpage}
\pagerange{\pageref{firstpage}--\pageref{lastpage}}
\maketitle

\begin{abstract} 
Outflows in starburst galaxies  
 driven by thermal-mechanical energy, 
  cosmic rays and their mix are investigated 
  with 1D and 2D hydrodynamic simulations.  
We show that these outflows 
  could reach a stationary state, 
  after which their hydrodynamic profiles 
  asymptotically approach 
  previous results obtained semi-analytically 
  for stationary outflow configurations.
The X-rays from the simulated outflows are computed, 
  and high-resolution synthetic spectra and 
  broadband light curves are constructed. 
The simulated outflows 
  driven by thermal mechanical pressure and CRs 
  have distinguishable spectral signatures, 
  in particular, 
  in the sequence of the keV K$\alpha$ lines 
  of various ions 
  and in the L-shell Fe emission complex. 
We demonstrate that broadband colour analysis in X-rays
  is a possible alternative means  
  to probe outflow driving mechanisms 
  for distant galaxies, 
  where observations may not be able  
  to provide sufficient photons 
  for 
  high-resolution spectroscopic analyses.  
  

\end{abstract}





\section{Introduction}

Large-scale outflows (winds) in galaxies 
  are often initiated by star-forming activities, 
  e.g. in the nearby starburst 
  galaxies Arp 220~\citep[e.g.][]{Varenius2016AA, Barcos2018ApJ}, M82~\citep[e.g.][]{Shopbell1998ApJ, Bland1988Natur} and NGC 253~\citep[e.g.][]{Walter2017ApJ, Krieger2019ApJ}, while more distant, younger galaxies are also known to host such phenomena~\citep{Frye2002ApJ, Rubin2014ApJ, Rupke2005ApJS-a, Arribas2014A&A}.
These outflows develop from the   
  confluence of stellar winds from young stars, 
  and energetic particles and gas from supernova events. 
Galactic outflows tend to have a bi-conical structure, 
  extending above and below the plane of their host galaxy,  
  following the path of least resistance through the interstellar environment
  \citep{Veilleux2005}. 
The structures of outflows, however, 
  vary among galaxies,  
  showing a variety of subtleties in morphology and dynamics. 
Velocities of galactic outflows 
  are mostly between $10-1000$ km s$^{-1}$~\citep{Cecil2002RMxAACS, Rupke2005ApJS-b, Rubin2014ApJ}. 
Outflow extents
  may reach a few tens of kpc~\citep{Veilleux2005, Bland-Hawthorn2007APSS, Bordoloi2011ApJ, Martin2013ApJ, Rubin2014ApJ, Bordoloi2016MNRAS},  
  often with X-ray emitting gas at lower altitudes bound 
  by a cap a few kpc from the starburst nucleus~\citep{Strickland2000AJ, Cecil2002ApJ, Cecil2002RMxAACS, Devine1999, Tsuru2007}. 
The gas temperature in some outflows may reach $10^7$K 
  \citep{McKeith1995A&A, Shopbell1998ApJ}. 
There is evidence of 
  multi-phase structure in galactic winds,  
  with cooler dense clumps, of temperature $10^2-10^4\ {\rm K}$ \citep{Strickland1997A&A, Lehnert1999ApJ}, 
  entrained within the hot gas 
   \citep[see][]{Heckman2003,Zhang2014ApJ, Walter2017ApJ, Wu2020}. 
The presence of this multi-phase structure  
   implies the coexistence 
   of material with very different structural and local thermal conditions. 
The diversity of the physical conditions within an outflow
  naturally gives rise to a multitude 
  of radiative processes, 
  and this complexity is not often present 
  in more uniform astrophysical systems. 
 
 Non-AGN associated outflows 
  are a consequence of star-forming activities in galaxies. 
  Such outflows carry away energy, gas and metals 
  from star-forming regions 
  to the circum-galactic medium of their host galaxy. 
This regulates the inflow of circum-galactic gas 
  and, hence, subsequently star formation activity, thus acting as a feedback mechanism alongside other processes operating within the interior of the host~\cite[e.g.][]{Owen2018MNRAS}.  
This inter-dependence between galactic outflows 
  and the star-forming activity of a galaxy 
  implies that the two processes co-evolve together  
  \citep[see][]{Mannucci2010}. 
Thus, the thermal and hydrodynamic (HD) properties 
  and the observational characteristics of galactic outflows 
  would show evolutionary trends 
  \citep[see][]{Sugahara2019ApJ}.   
This is consistent with the finding that galactic properties and sizes 
  vary over redshift   
  \citep[with younger galaxies at higher redshift being bluer and smaller than 
  those at $z\sim0$, see][]{Madau1996, Dickinson2003}.



Outflows driven 
  by thermal mechanical pressure, radiation, and cosmic rays (CRs),  
  were investigated by \citealt{Yu2020} (hereafter~\citetalias{Yu2020})  
  using a phenomenological HD model  
  \citep[see also][]{Chevalier1985Nat, thompson2015, Sharma2013, Ipavich1975, Samui2010}.   
Outflows predominantly driven by thermal mechanical pressure    
  were found to be the hottest 
  and have the highest velocities. 
Radiation-driven outflows 
  are unable to attain similar velocities for realistic opacities.  
These tend to be cooler and denser, 
  and are prone to cooling instability.
 CR-driven outflows have velocities which 
  fall between thermal-mechanically driven and radiatively driven cases.  
They could be very extended in altitude~\citep[see, e.g.][]{Jacob2018}, 
  when their entrained magnetic field is sufficiently strong 
  to interact with CRs.  

Although analytical studies, 
  e.g. \cite{Chevalier1985Nat} and \citetalias{Yu2020}, 
  provide some useful insights, 
  they have their limitations 
  \citep[cf. the recent numerical simulations of ][]{Tanner2020ApJ}, 
  e.g. they cannot describe the time-dependent development of an outflow,  
  or model the transition of an outflow from a transient to a stationary state. 
They are also unable to determine 
  if an outflow even has a stationary state. 
An example of such would be an outflow from a galaxy 
  with a very massive dark-matter halo.  
Outflow gas would be trapped by its deep gravitational well, 
  being unable to escape into intergalactic space. 
The metal-enriched gas trapped under the gravity of the dark-matter halo 
  would return back to the host galaxy, 
  fueling subsequent episodes of star formation 
  \citep[see e.g.][for various aspects of the galactic recycling process]{Fox2017ASSL}. 
Another example is the fragmentation of a flow 
  subject to strong radiative cooling.  
This induces instabilities 
  and leads to the development of multi-phase gas components in the flow. 
The phase transition between the components is regulated by 
  thermal mechanical (e.g. shocks) and ionisation processes,   
  which compete with each other globally and locally 
  \citep[see e.g.][]{Hoopes2003}, 
  hence contributing 
  to the regulation of the HD energy and momentum budget.  
 
This paper presents a time-dependent study 
  of galactic outflows using HD simulations.  
  It is an extension of 
  the phenomenological study of \citetalias{Yu2020},   
  with a focus on outflows driven by thermal mechanical pressure, 
  CRs and their mix.    
\S~\ref{sec:method}   
  presents the HD formulation  
  and its numerical implementation 
  using a \verb|FLASH|-based code. 
\S~\ref{sec:results_HD}  
  shows the results of our numerical simulations 
  of galactic outflows with the considered driving mechanisms,  
  and \S~\ref{sec:XR_emission}
  shows the X-ray emission computed from the simulated outflows. 
Astrophysical implications are discussed in \S~\ref{sec:discussion},  
  and a conclusion is presented in \S~\ref{sec:conclusion}. 

\section{Time-dependent Model}
\label{sec:method}

The outflows are comprised of 
  two coupled fluid components: 
  a hot ionised thermal gas, and CRs. 
The flows are inviscid and non-turbulent. 
CRs are treated as a relativistic non-thermal fluid 
  with pressure but negligible bulk kinetic energy  
  \citep[cf. the approach adopted by][]{Ipavich1975, Breitschwerdt1991AA}.  
The outflows are driven by either  
  thermal mechanical pressure 
  (associated with the thermal content of the gas), CRs 
  or their mix. 
Radiative cooling is included in the simulations, unless otherwise stated.  
A spherically symmetric geometry 
  is assumed in the 1-dimensional (1D) simulations, 
  and a cylindrical geometry is assumed 
  in the 2-dimensional (2D) simulations. 

The HD formulation 
  is the same as in \citetalias{Yu2020}, 
  but with full consideration 
  of the evolutionary and time-dependent properties 
  of the flows. 
This formulation is valid, 
  provided that the ram pressure of a galactic outflow 
  is sufficiently larger than 
  the pressure of the ambient gas.  
(Note that the radial profiles and 2D plots  
   shown later in this paper extend only to 10 kpc,  
   which is smaller than the virial radius of their host galaxy,   
   to ensure this constraint is met.)  
Unless otherwise stated, 
  we adopt the reference model parameters listed
  in Table~\ref{tab:param}.  
These parameters are appropriate for star-forming galaxies 
  resembling the nearby starburst galaxy M82. 


\subsection{Hydrodynamic formulation}
\label{sec:HD} 


The HD equations describing the outflows are:
\begin{gather}
\label{eq:HD_mass}
\frac{\partial\rho}{\partial t}+\nabla\cdot\left(\rho{\boldsymbol v}\right)=q\ ,\\
\label{eq:HD_momentum}
\frac{\partial\rho\textbf{v}}{\partial t}
  +\nabla\cdot\left(\rho{\boldsymbol v}{\boldsymbol v}\right)=\rho {\boldsymbol f}-\nabla P-\nabla P_{\rm C}\ ,\\
\label{eq:HD_energy}
\frac{\partial\rho E}{\partial t}
 +\nabla\cdot\left[\left(\rho E+P\right){\boldsymbol v}\right]
  =Q-C+\rho{\boldsymbol v}\cdot{\boldsymbol f}+I\ ,\\
\label{eq:HD_enec}
\frac{\partial\rho_{\rm C}E_{\rm C}}{\partial t}+\nabla\cdot\left[\left(\rho_{\rm C} E_{\rm C}+P_{\rm C}\right){{\boldsymbol v}_{\rm C}}\right]=Q_{\rm C}-I\ , 
\end{gather}
Here, 
  $\rho$ is density,
 ${\boldsymbol v}$ is velocity, 
 $P$ is pressure and 
 $E$ is total energy per unit mass 
 for the fluid component.  
The corresponding quantities  
 for the CR component are 
$\rho_{\rm C}$, ${\boldsymbol v}_{\rm C}$, $P_{\rm C}$ and $E_{\rm C}$.  
${\boldsymbol f} (= {\boldsymbol f}_{\rm rad} + {\boldsymbol f}_{\rm grav}$) is  the external force term, 
  which is the sum of radiative and gravitational forces 
  (see \S~\ref{sec:forces}).  
$C$ is the radiative cooling rate of the thermal gas, $I$ is the energy transfer rate from the CR to the thermal component of the wind, 
  and $q$, $Q$ and $Q_{\rm C}$ 
  are the injection rates of mass, 
  energy and CR energy respectively (see \S~\ref{sec:injection_terms} for details). 
Setting the partial time derivatives in 
  Equations~\ref{eq:HD_mass} to~\ref{eq:HD_enec} 
  to zero gives the set of HD equations for stationary-state flows. Solutions to these were obtained by \citetalias{Yu2020}, providing a model describing galactic outflows after they have reached a stationary state.
Here, we consider more general time-dependent flows with specific driving mechanisms and initial conditions,  
  using numerical HD simulations.  
  




\subsubsection{Cosmic rays}
\label{sec:crs}

The CRs are treated as a relativistic fluid with pressure. 
The bulk kinetic energy of the CRs,
which is negligible compared to the CR pressure,
is assumed to have 
  no direct mechanical interaction 
  with the gas component of the outflow.  
The rate of energy transfer from the CRs to the gas, 
  which is magnetised and is assumed to be in local thermal equilibrium,  
  is given by 
\begin{equation}
 I = -({\boldsymbol v} + {\boldsymbol v}_{\rm A})\cdot \nabla P_{\rm C} \ ,
\end{equation}
  \citep[see][]{Ipavich1975, Breitschwerdt1991AA}, 
  where ${\boldsymbol{v}}_{\rm A}$ is the Alfv\'en velocity. 
Unlike AGN jets, 
  galactic wind outflows are not magnetically collimated, 
  and the magnetic fields are carried along with gas flows.  
The magnetic fields are therefore tangled.      
As such, we may take an effective field strength 
  to derive the magnitude of the Alfv\'en velocity.  
Without losing generality, 
  the global direction of the Alfv\'en velocity vector 
  is assumed to be in the same direction as the flow velocity, i.e. 
${\boldsymbol{v}}_{\rm A}=\langle|B(r)|\rangle{\boldsymbol{\hat{r}}}/\sqrt{4\pi \rho}$, where
\begin{equation}
\label{eq:B_field}
\langle|B(r)|\rangle=\begin{cases}
\frac{B_0 r}{r_{\rm sb}}&\text{if }r<r_{\rm sb}\\
\frac{B_0{r_{\rm sb}}^2}{r^2}&\text{if }r\geq r_{\rm sb}
\end{cases}\ ,
\end{equation}
   as in \citetalias{Yu2020}, 
   where $r = |{\boldsymbol r}|$ is the radial position along the outflow, and $\hat{\boldsymbol r} = {\boldsymbol r}/|{\boldsymbol r}|$. 

The radius of the starburst region 
  is set to be $r_{\rm sb}=250\ \rm pc$, 
  a value similar to that estimated for the M82 starburst core \citep[see][]{deGrijs2001}. 
The magnetic field strength of the starburst region 
  is set to $B_0=50\ \mu\rm G$ in the fiducial case.  
This strength is comparable 
   to estimates of M82 \citep{Klein1988}.  
The magnetic field here  
  is not self-consistently 
  derived from CR streaming and HD, 
  but for the purposes of this work, 
  this treatment will give 
  an acceptable rate for the transfer 
  of energy from the CR fluid to the gas.  
A more complicated numerical 
  magneto-hydrodynamic treatment 
  does not necessarily give more reliable results, 
  when the exact interactions 
  between CRs, magnetic fields 
  (including their seeding) 
  and gas 
  (which is clumpy, not fully ionised 
  and out of local thermal equilibrium) 
  are uncertain.  
Our previous calculations \citepalias{Yu2020}, 
  which were conducted under a stationary condition, 
  demonstrated 
  that outflows with $B = 0 \mu{\rm G}$ 
  (where CR heating cannot be facilitated) 
  would yield temperatures several times lower 
  than outflows with $B\sim 100 \mu {\rm G}$ 
  (where CR heating can contribute substantially). 
Similar effects are expected 
  in outflows, 
  where the stationary condition is relaxed.


\begin{table}
\centering
\begin{tabular}{*{3}{c}}
\midrule
Parameter & Value & Reference \\
\midrule
$r_{\rm sb}$ & $250\ \rm pc$ & \cite{deGrijs2001} \\
$\mathcal{R}_{\rm SF}$ & $10\ M_\odot/\rm yr$ & \cite{deGrijs2001} \\
$\dot{E}$ & $7\times10^{42}\ \rm erg/s$ & \cite{Veilleux2005}  \\
$\dot{M}$ & $2.6\ M_\odot/\rm yr$ & \cite{Veilleux2005} \\
$\kappa$\textsuperscript{\textit{a}} & $\sim10^4\ \rm cm^2/g$ & \cite{Li2001} \\
$B_0$\textsuperscript{\textit{b}} & $50\ \mu \rm G$ & \cite{Klein1988} \\
$M_{\rm tot}$\textsuperscript{\textit{c}} & $5.54\times 10^{11}\ M_\odot$ & \cite{Oehm2017} \\
$R_{\rm s}$\textsuperscript{\textit{c}} & $14.7\ \rm kpc$ & \cite{Oehm2017} \\
$R_{\rm vir}$\textsuperscript{\textit{c}} & $164\ \rm kpc$ & \cite{Oehm2017} \\
\midrule 
\end{tabular}
\caption{A list of reference parameters 
 for the galactic outflows in the fiducial model of our HD simulations. 
 These parameters are appropriate for 
   starburst galaxies resembling the nearby starburst galaxy M82. 
Notes: \newline
\textsuperscript{\textit{a}} \(\kappa\) is the mean opacity of the wind, averaged over all frequencies. \newline
\textsuperscript{\textit{b}} \(B_0\) is the maximum galactic magnetic field strength (external to the wind). \newline
\textsuperscript{\textit{c}} \(M_{\rm tot}\), \(R_{\rm s}\) and \(R_{\rm vir}\) are the parameters for the \citealt{Navarro1996} DM profile.}
\label{tab:param}
\end{table}

\subsubsection{External forces}
\label{sec:forces} 

In equation~\ref{eq:HD_energy}, 
  ${\boldsymbol f}$ represents the external forces acting on the gas, which is the combination of 
  the radiative force and the gravitational force. 
The radiative force is provided by the radiation from the starburst core.  
It is proportional to the gas density $\rho$ 
  and the effective opacity of the gas, $\kappa$, weighted across all wavelengths. 
It also depends on the radiation field, 
  and is hence determined 
  by the local radiation energy density, 
  which, at the location $r$, is given by 
  $L/4\pi r^2c$, where $c$ is the speed of light.  
In our calculations,   
  we take $\kappa=10^4\rm\ cm^2\ g^{-1}$ \citep{Li2001}, 
  which corresponds to a strong radiative driving scenario. 
For radiation uniformly emitted from the starburst core,  
  the luminosity, as seen by the gas at $r$,
  can be parametrised as  
\begin{equation}
\label{eq:L}
  L\left(r\right)=L_{0}\begin{cases} 1&\text{if }r\geq r_{\rm sb}\\
\frac{r^3}{{r_{\rm sb}}^3}&\text{if }r<r_{\rm sb}\end{cases}\ . 
\end{equation}
This implicitly assumes 
  that the decrease in luminosity over distance 
  is predominantly caused by geometrical dilution of the radiation field, 
  instead of by absorptive attenuation  
  through its interaction with the outflowing gas. 
The luminosity of the radiation generated by the starburst core, $L_0$,
  specifies the radiative power injected 
  into the system (see also \S~\ref{sec:injection_terms}). 

The gravitational force is 
  provided by the dark matter (DM) halo of the galaxy, 
  which takes the form of a Navarro-Frenk-White profile: 
\begin{equation}
\label{eq:M}
    M\left(r\right)
    =\frac{M_{\rm tot}\xi\left(r\right)}{\xi\left(R_{\rm vir}\right)} \ , 
\end{equation}
where
\begin{equation}
   \xi\left(r\right)=\ln\frac{R_{\rm s}+r}{R_{\rm s}}-\frac{r}{R_{\rm s}+r} \ , 
\end{equation}
\citep{Navarro1996}. 
  Here  
  $M_{\rm tot}=M\left(R_{\rm vir}\right)$ is the total halo mass, 
  $R_{\rm vir}$ is the virial radius of the halo, 
  and $R_{\rm s}$ is the halo scale radius. 
The fiducial values adopted in this study are 
  $M_{\rm tot}=5.54\times10^{11}\ {\rm M}_\odot$, 
  $R_{\rm vir}=164\ \rm kpc$ 
  and $R_{\rm s}=14.7\ \rm kpc$, from \cite{Oehm2017}. 
These values are appropriate for the DM halo of a galaxy similar to M82. 
The total external force, accounting for radiation and gravity,
  is therefore  
\begin{align}
\label{eq:external}
{\boldsymbol f} 
   &= {\boldsymbol f}_{\rm rad}+ {\boldsymbol f}_{\rm grav} 
   =\frac{\kappa L\left(r\right)}{4\pi {r}^2c }\;\!\hat{\boldsymbol r}-\frac{GM\left(r\right)}{{r}^2}\;\!\hat{\boldsymbol r} \  .
\end{align}

\subsubsection{Source terms}
\label{sec:injection_terms}  

The source terms in the HD equations are given by
\begin{equation}
\label{eq:density}
q = \frac{3\dot{M}}{4\pi {r_{\rm sb}}^3} \begin{dcases}
1 &\text{if }r< r_{\rm sb} \ , \\
0&\text{if }r\geq r_{\rm sb} \ ,
\end{dcases} \\
\end{equation}
for mass, 
\begin{equation}
Q = \frac{3\dot{E}}{4\pi {r_{\rm sb}}^3} \begin{dcases}
1 &\text{if }r< r_{\rm sb} \ , \\
0&\text{if }r\geq r_{\rm sb} \ ,
\end{dcases} \\
\end{equation}
for energy, and 
\begin{equation}
Q_{\rm C} = \frac{3\dot{E}_{\rm C}}{4\pi {r_{\rm sb}}^3} \begin{dcases}
1 &\text{if }r< r_{\rm sb} \ , \\
0&\text{if }r\geq r_{\rm sb} \ ,
\end{dcases} \\
\end{equation}
for CR energy,  
  where $\dot{M}$ is the total mass injection rate, 
  and $\dot{E}$ and $\dot{E}_{\rm C}$ are respectively 
  the energy injection rates into the thermal gas and CR fluid 
  across the entire starburst region. 
Unless stated otherwise, matter and energy are injected at constant rates,   
  uniformly throughout the starburst region. 

The mass injection rate, $\dot{M}$, and the total energy injection rate, $\dot{E}_{\rm tot} = \dot{E} + \dot{E}_{\rm C}+L_0$, are parametrised by the star formation rate (SFR) $\mathcal{R}_{\rm SF}$, and are given by 
\begin{equation}
    \dot{M} = 0.26 ~ \mathcal{R}_{\rm SF} \; {\rm M}_\odot\ \rm yr^{-1} \ ,
\end{equation}
where 0.26 is the estimated fraction of mass lost to stellar winds and supernovae (SNe) (computed using Starburst99, if adopting solar metallicity -- see \citealt{Leitherer1999, Veilleux2005}), and
\begin{equation}
    \dot{E}_{\rm tot}=7\times10^{41}~\left(\frac{\mathcal{R}_{\rm SF}}{{\rm M}_\odot\ \rm yr^{-1}}\right) ~{\rm erg}\;\!{\rm s}^{-1}
\end{equation}
\citep{Veilleux2005, Leitherer1999}, with values scaled from M82, for which $\mathcal{R}_{\rm SF} \approx 10 ~{\rm M}_\odot\ \rm yr^{-1}$~\citep{deGrijs2001}. 90\% of the total injected energy is radiated away \citep{Leitherer1999, Thornton1998}, and only the remaining 10\% is available to drive the system (via CRs or thermal pressure). We take the CR and thermal powers to be equal in our fiducial model, i.e.
\begin{align}
\label{eq:E}
\dot{E} &= \alpha \dot{E}_{\rm tot} \ ,\\
\label{eq:Ec}
\dot{E}_{\rm C} &= \alpha_{\rm C} {\dot{E}_{\rm tot}} \ ,\\
\label{eq:EL}
L_{0} &= \alpha_{\rm L} {\dot{E}_{\rm tot}} \ ,
\end{align}
where $\alpha = 0.05$, $\alpha_{\rm C} = 0.05$, and $\alpha_{\rm L} = 0.9$. 
We vary and investigate these driving efficiencies in \S~\ref{sec:CR}.


\subsubsection{Cooling}
\label{sec:radiative_cooling}

Cooling processes play an important role 
  in regulating the thermal properties and HD of galactic outflows 
  and, hence, their observational characteristics 
  \citep[see e.g.][]{Silich2004}. 
A self-consistent treatment of radiative heating and cooling 
  in the presence of magnetic fields and CRs 
  is non-trivial and computationally demanding. 
For the purposes of this work, 
a simplified treatment of a cooling function is sufficient to give acceptable qualitative results to investigate the impacts of the three different driving mechanisms  
  of galactic outflows. 
A cooling function generally 
  depends on the local properties of the flows, 
  such as the thermal temperature $T$ and gas density $\rho$. 
The \verb|CLOUDY| code\footnote{\url{http://www.nublado.org/}} \citep{Ferland2017} 
  is used to compute the cooling function, $\Lambda\left(T\right)$ 
  (see also \citealt{Sutherland1993ApJS}),  
   giving a cooling rate 
   $C = \rho^2 \Lambda\left(T\right) / \mu^2$  
   (see equation~\ref{eq:HD_energy}),  
   where $\mu=1.4\ m_{\rm H}$ is the mean molecular mass, 
   i.e the same value 
   as that adopted in \cite{Veilleux2005}.  \\


\subsection{Code implementation and treatment of cosmic rays} 
\label{sec:FLASH}

\verb|FLASH|\footnote{\url{http://flash.uchicago.edu/}} version 4~\citep{Fryxell2000, Dubey2008}, a Eulerian grid code, 
  is used as the numerical solver for the HD 
  equations \ref{eq:HD_mass} to \ref{eq:HD_enec}.  
The CRs are treated as a single fluid, i.e.  
  the CR species, their energy spectra 
  and the CR cooling/heating processes 
  associated with individual species are not considered explicitly. 
The energy evolution of the CR fluid is determined by
$E_{\rm C}$, where the equation of state, $P_{\rm C}=\left(\gamma_{\rm C}-1\right)\rho_{\rm C}E_{\rm C}$, is adopted with an adiabatic index $\gamma_{\rm C}=4/3$. The thermal gas pressure is computed using the equation of state $P=\left(\gamma-1\right)\rho E$ where, this time, the adiabatic index is $\gamma=5/3$. All source terms are implemented with a 2nd order central difference method.

The initial values of  
  $\rho=10^{-28}\rm\ g\ cm^{-3}$, $T=10\rm\ K$, $P_{\rm C}=P$ and $\boldsymbol{v}=0\rm\ km\ s^{-1}$ 
  are used throughout the grid in all the simulations, 
  with other necessary initial HD quantities being derived from these. 
The 1D simulations 
  have spherical symmetry 
  and an extent of 10~kpc, 
  with a resolution of about 4.88~pc. 
The 2D simulations have cylindrical symmetry
 in a square simulation grid 
 with 10 kpc on each side, 
 and a resolution of about 19.53~pc.
Uniform grid is used in all simulations, and the "outflow" boundary conditions are set at the edge of the simulation domain, except in cases where "reflecting" boundary conditions (e.g. between quadrants in 2D cylindrical simulations) can make use of the symmetry of the model to avoid unnecessary duplication of computations in equivalent simulation regions.

\subsection{Numerical simulations}
\label{sec:spec}

\begin{table*}
\centering
\setlength{\tabcolsep}{5.5pt}
\begin{tabular}{*{10}{c}}
Run ID & 1D/2D & Cooling & $\alpha$ & $\alpha_{\rm C}$ & $\alpha_{\rm L}$ & $\mathcal{R}_{\rm SF}(t)$ & $R_{\rm A}$ & HD section(s) & X-ray section(s) \\
\midrule
\verb|Sim1C| & 1D & \ding{51} & 0.05 & 0.05 & 0.9 & $\mathcal{R}_{\rm SF}$ & 1 & \ref{sec:cool} & \ref{sec:XR_spectroscopy}, \ref{sec:XR_broadband} \\
\verb|Sim1c| & 1D & \ding{55} & 0.05 & 0.05 & 0.9 & $\mathcal{R}_{\rm SF}$ & 1 & \ref{sec:cool} & \ref{sec:XR_cooling_evolution}, \ref{sec:XR_broadband} \\
\verb|Sim2C| & 2D & \ding{51} & 0.05 & 0.05 & 0.9 & $\mathcal{R}_{\rm SF}$ & 1 & \ref{sec:cool} & N/A \\
\verb|Sim2c| & 2D & \ding{55} & 0.05 & 0.05 & 0.9 & $\mathcal{R}_{\rm SF}$ & 1 & \ref{sec:cool} & N/A \\
\midrule
\verb|Sim1CR| & 1D & \ding{51} & 0.025 & 0.075 & 0.9 & $\mathcal{R}_{\rm SF}$ & 1 & \ref{sec:CR} & \ref{sec:XR_driving}, \ref{sec:XR_redshift}, \ref{sec:XR_broadband} \\
\verb|Sim1th| & 1D & \ding{51} & 0.075 & 0.025 & 0.9 & $\mathcal{R}_{\rm SF}$ & 1 & \ref{sec:CR} & \ref{sec:XR_driving}, \ref{sec:XR_redshift}, \ref{sec:XR_broadband} \\
\midrule
\verb|Sim1P| & 1D & \ding{51} & 0.05 & 0.05 & 0.9 & $\mathcal{R}_{\rm SF}+\rm noise$ & 1 & \ref{sec:noise} & N/A \\
\verb|Sim2P| & 2D & \ding{51} & 0.05 & 0.05 & 0.9 & $\mathcal{R}_{\rm SF}+\rm noise$ & 1 & \ref{sec:noise} & N/A \\
\verb|Sim1S| & 1D & \ding{51} & 0.05 & 0.05 & 0.9 & $\mathcal{R}_{\rm SF}(1+\rm{sin}(2\pi t/\tau))$ & 1 & \ref{sec:SFR} & \ref{sec:XR_broadband} \\
\midrule
\verb|Sim2C2| & 2D & \ding{51} & 0.05 & 0.05 & 0.9 & $\mathcal{R}_{\rm SF}$ & 2 & \ref{sec:shape} & N/A \\
\verb|Sim2C5| & 2D & \ding{51} & 0.05 & 0.05 & 0.9 & $\mathcal{R}_{\rm SF}$ & 5 & \ref{sec:shape} & N/A \\
\verb|Sim2C10| & 2D & \ding{51} & 0.05 & 0.05 & 0.9 & $\mathcal{R}_{\rm SF}$ & 10 & \ref{sec:shape} & N/A \\
\midrule 
\end{tabular}
\caption{Configuration of the four sets of simulation runs, based on the parameter values in Table~\ref{tab:param}. The first set assesses the impact of radiative cooling and the level of agreement between 1D and 2D simulations. The second set investigates the energy partition of different driving mechanisms in the outflow. The third set models variations in the SFR, and the fourth set considers variations in the geometry of starburst region. The corresponding discussion sections in this paper are also indicated.}
\label{tab:sim}
\end{table*}

We perform numerical simulations to investigate the evolution and characteristics of outflows emerging from star-forming galaxies.
They are appropriate for outflows in starburst galaxies similar to M82. The fiducial model in this study adopts the 
 parameters listed in Table~\ref{tab:param}, and 
four sets of simulations are conducted 
  to investigate different scenarios, 
  whose characteristics are summarised in Table~\ref{tab:sim}.

The first set of simulations investigates 
  the effects of radiative cooling, 
  and the consistency between 1D and 2D simulations. 
\verb|Sim1C| is the 1D reference case 
  (with parameters in Table~\ref{tab:param}). 
\verb|Sim1c| adopts the same parameter choices, but 
  with no radiative cooling.
\verb|Sim2C| and \verb|Sim2c| 
  are the corresponding cases in a 2D configuration. 
 
The second set of simulations investigates 
  the relative efficiency of thermal mechanical pressure and CRs  
  in driving the outflow. 
For realistic opacities, 
  radiation is inefficient in driving an outflow  
  when compared to thermal mechanical pressure and CRs.   
We set $\alpha_{\rm L}=0.9$ \citep{Leitherer1999, Thornton1998} 
  in this set of simulations, 
  so that we can discern 
  the partition between thermal mechanical and CR drivings 
  with more clarity. 
In \verb|Sim1C|, the reference case, 
  the fractions of energy supplied to the thermal and CR components of the outflow are equal, i.e. $\alpha_{\rm C}=0.05$ and $\alpha=0.05$. The injection of energy supplied to CRs dominates in \verb|Sim1CR| where $\alpha_{\rm C}=0.075$ and $\alpha=0.025$, while \verb|Sim1th| is thermally dominated with $\alpha_{\rm C}=0.025$ and $\alpha=0.075$.
 


The third set of simulations investigates 
  the effects of the variations in $\mathcal{R}_{\rm SF}$.   
In \verb|Sim1P| and \verb|Sim2P|,    
  we assign the variations as stochastic fluctuations, 
  which is a Poisson process caused by SN events.  
A SN event rate $\dot{N}_{\rm SN}=0.02~\mathcal{R}_{\rm SF}\rm\;yr^{-1}$  
  \citep[][]{deGrijs2001, Veilleux2005} 
  is adopted, 
  and the event rate in each time step of the simulation 
  is determined following a Poisson distribution. 
In \verb|Sim1S| we consider multiple episodes of star-forming activities. 
Without losing generality, 
 we consider periodic star formation episodes. 
The strengths of the episodes are roughly equal 
  and a sinusoidal model is adopted for the SFR:  
\begin{equation}
    \mathcal{R}_{\rm SF}(t) = \mathcal{R}_{\rm SF, 0} \left(1 + 0.99\;\sin \left[ \frac{2\pi t}{\tau}\right]\right) \ ,
\label{eq:sin}
\end{equation}
 with $\tau = 100\;\!{\rm Myr}$ 
  and $\mathcal{R}_{\rm SF, 0} = 10 \;\! {\rm M}_{\odot}\;\!{\rm yr}^{-1}$. 
This suffices to mimic  
  successive alternation of quenching and rejuvenation 
  of star formation in the galaxy. 

The last set of simulations investigates 
  the effects of the starburst region's geometry. 
Without losing generality, 
  we adopt a simple geometrical model 
  in terms of the aspect ratio of a starburst core. 
The starburst core is an ellipsoid,  
  with a height  $z_{\rm sb}$ 
  and radius $r_{\rm sb}$, 
  which can be parametrised by the ratio 
  $R_{\rm A}=r_{\rm sb}/z_{\rm sb}$. 
In the simulations, 
  three values of $R_{\rm A}$ 
  are considered: 2, 5 and 10. 
The volume of the ellipsoid $V$ is conserved in all cases, with $V = {4\pi}r_{\rm sb}^3/3$, where $r_{\rm sb}$ retains its fiducial value.
\section{Hydrodynamics of outflows}
\label{sec:results_HD}

\subsection{General characteristics}
\label{sec:general}

\begin{figure*}
\includegraphics[width=0.9\textwidth]{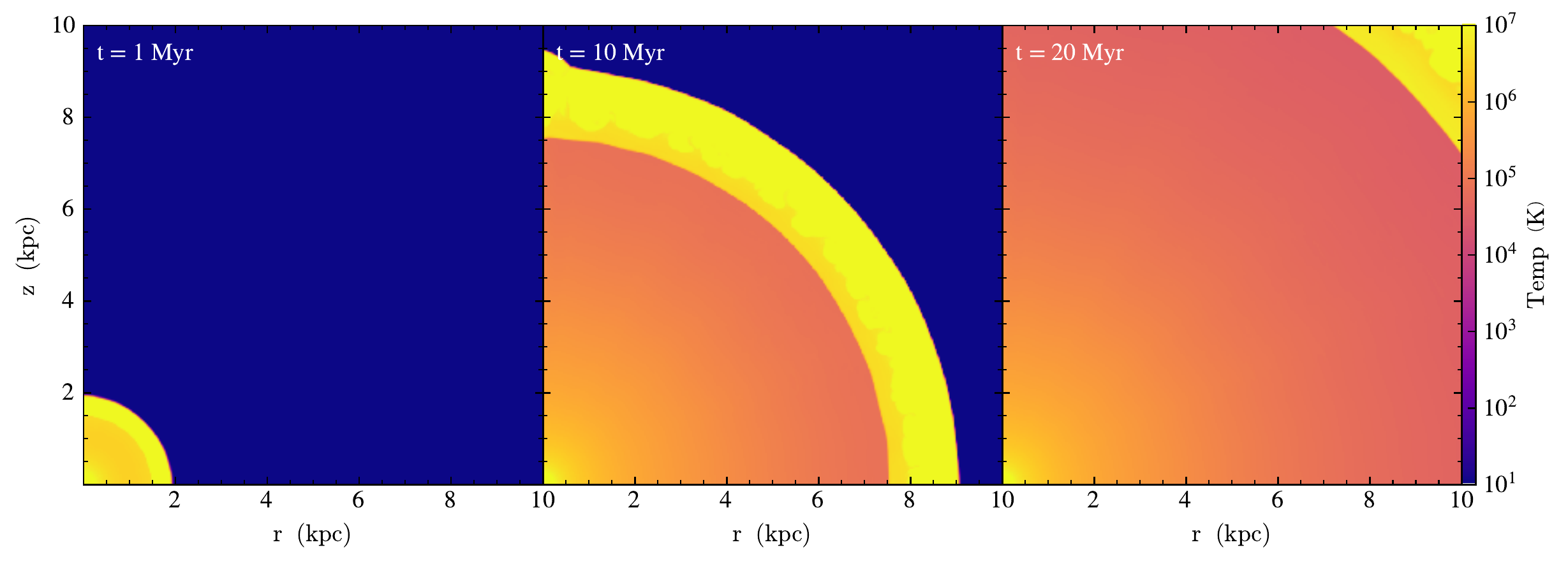}
\caption{2D temperature maps for \texttt{Sim2C} (the fiducial case) after 1 Myr, 10 Myr and 20 Myr, demonstrating the evolution of the simulated outflow. While present, the carbuncle instability is broadly suppressed by use of a split PPM solver.}
\label{fig:HD2D_Sim2C}
\end{figure*}

\begin{figure*}
\includegraphics[width=0.9\textwidth]{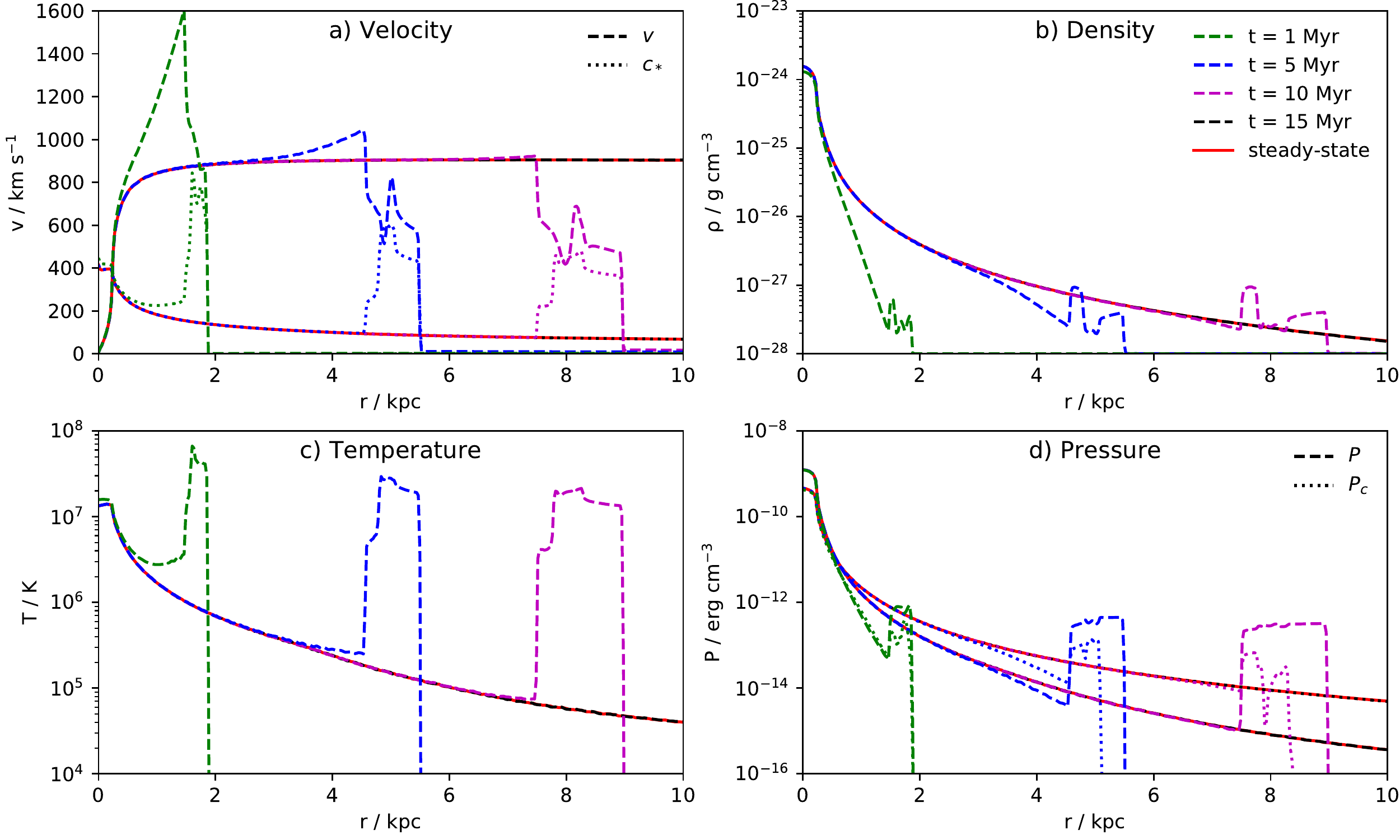}
\caption{HD profiles for an outflow developing to a stationary state, with snapshots of \texttt{Sim2C} (the fiducial case) shown at 1, 5, 10, 15 Myr, and the stationary-state solution at the end of the simulation. Outflow velocity and effective sound speed are plotted in panel \textit{a}; gas density is plotted in panel \textit{b}; gas temperature is plotted in panel \textit{c}; thermal gas and CR pressures are shown in panel \textit{d}.
The density profile of the stationary-state outflow in panel \textit{b} follows $\rho\propto r^{-2}$, while the pressure profiles in panel \textit{d} satisfy $P\propto\rho^\gamma$ and $P_{\rm C}\propto\rho^{\gamma_{\rm C}}$.
The spherical symmetry of the system means that the profile should not vary with polar angle, however we plot the profiles at $\theta=\pi/4$ in all cases to avoid any numerical artifacts that may emerge (e.g. from the carbuncle instability).} 
\label{fig:HD1D_Sim2C}
\end{figure*}

Figure~\ref{fig:HD2D_Sim2C} shows the 2D temperature map 
  in simulated galactic outflows 
  at three different evolutionary stages for the fiducial case, \verb|Sim2C|.  
 Figure~\ref{fig:HD1D_Sim2C} 
 shows the corresponding  
 radial velocity, sound speed, density, 
 temperature and pressure profiles at four different evolutionary stages,
 against the stationary-state profiles 
 obtained at the end of the simulations.
As shown, 
  the stationary state is reached by $\sim$ 15 Myr 
  in this case. 
The right-most panel 
  in Figure~\ref{fig:HD2D_Sim2C} 
  shows the temperature profile 
  of the outflow at 20~Myr, 
  which is essentially 
  the stationary-state temperature profile. 
The asymptotic stationary-state profiles of the flows 
  shown here 
  are in good agreement with those obtained by calculations 
  using the analytic approach of \citetalias{Yu2020} (see Appendix~\ref{sec:comparison_analytic}), 
  if adopting the same set of system parameters.

The interstellar gas is initially uniform in density and temperature. 
The injection of material and energy from the starburst region
  exerts a pressure force onto the surrounding interstellar medium (ISM), 
  pushing gas outwards.    
The material leaves the starburst region, accelerates, and 
  a hot ``bubble'' of gas begins to develop. 
As this bubble expands, an outflow is launched,  
  with its maximum value $v_{\rm max}$ 
  greatly exceeding the velocity expected for a stationary-state flow.   
Its front quickly acquires a supersonic speed 
  (cf. the flow velocity and the sound speed\footnote{The effective sound speed $c_*$ 
  is given by
\begin{equation}
{c_*}^2=\frac{\gamma P}{\rho}+\frac{\gamma_{\rm C}P_{\rm C}}{\rho}\frac{\left(2\boldsymbol{v}+\boldsymbol{v}_{\rm A}\right)\cdot \left(\boldsymbol{v}-\left(\gamma-1\right)\boldsymbol{v}_{\rm A}\right)}{2\boldsymbol{v}\cdot\left(\boldsymbol{v}+\boldsymbol{v}_{\rm A}\right)}   
\end{equation}
\citep{Ipavich1975,Breitschwerdt1991AA}.} profiles 
  in panel \textit{a} of Figure~\ref{fig:HD1D_Sim2C}), 
  and a shock is formed 
  in the interface between the outflow and the ISM.  

As the hot bubble expands and the outflow proceeds, 
  the shock propagates outward 
  and a shock transitional layer is formed.    
The cooling time of the material in the shock transition layer 
  is longer than the dynamical time of shock propagation, 
  and the shock is approximately locally adiabatic. 
This is indicated in panel \textit{c} of Figure~\ref{fig:HD1D_Sim2C}, 
  as the shock jump temperature 
  does not decrease significantly, 
  and the thickness of the shock transitional layer 
  grows with the bubble's expansion, i.e. with the development of the outflow.  
Behind the shock transitional layer, 
  the post-shock flow settles asymptotically 
  to attain its stationary-state profile, 
  on a time scale $\sim 10 - 15$~Myr 
  (see panels \textit{b} and \textit{d} of Figure~\ref{fig:HD1D_Sim2C}).

When the shock sweeps through the ISM, 
  it produces a hot, dense shell of compressed gas.    
The swept-up material accumulates inertia 
  and thus slows down the shock propagation.  
By $\sim 15$~Myr,  
  $v_{\rm max}$ becomes comparable to the stationary-state ``wind'' outflow velocity 
   (\textit{a} of Figure~\ref{fig:HD1D_Sim2C}). 
At this stage, after the initiation of the material and energy injection,  
  the entire post-shock outflow behind the shock transitional layer 
  settles almost immediately into its stationary-state configuration.  
This is in contrast to post-shock flow 
  in the earlier stages in the development, 
  where $v_{\rm max}\gg v_{\rm st}$ 
  and a substantial amount of time 
  is required for the flow to settle into its corresponding stationary-state profile.

The timescale for the system to reach stationary state 
  is comparable to the global dynamical timescale of the outflow, 
  which may be expressed as 
  $t_{\rm out}=r/v_{\rm st}(\infty)$.   
Such time scales are much shorter than 
  the duration of a starburst episode (expected to last $\sim 100$~Myr) 
  of a starburst galaxy similar to M82. 
Thus, we may conclude 
  from this simulation 
  that the outflow would quickly settle into a stationary state 
  (if the conditions allow such a configuration to be reached), 
  implying that most of the outflows observed in the starburst galaxies 
  would have already settled into their stationary-state configurations.

While the outflow beyond the starburst region is supersonic, 
  the flow within the starburst region is subsonic in our model   
  (where $r<r_{\rm sb}$, for $r_{\rm sb} = 250~{\rm pc}$).  
With the subsonic flow in the region 
  and the flow velocity set to be $|\boldsymbol{v}|=0$ at $r=0$,  
  the velocity of the outflowing gas 
  would need to increase very substantially 
  in the starburst region and 
  through its interface with the outflow beyond the region.  
Note that the injection terms $q$, $Q$ and $Q_{\rm C}$ 
  adopted in the model  
  resemble a step-function, 
  which results in a very strong pressure gradient across 
  the boundary region $r=r_{\rm sb}$. 
With these distinct boundary conditions, 
  the acceleration in the starburst region 
  and through its boundary could be artificially extreme.    
The increase in flow velocity would be more moderate through this point 
  if a smoother boundary to the injection region is considered, 
  as shown in some studies \citep[e.g.][]{Owen2019a}. 
 Provided that 
   the total amount of material and energy injected 
   in the flow beyond the starburst region is properly accounted for,  
   the detailed treatment 
   of the micro-physics in the transitional boundary interface zone 
   has inconsequential global impacts 
   on the macroscopic HD structures and thermal properties 
   of a galactic outflow.

\subsubsection{Effects of cooling}
\label{sec:cool}

\begin{figure*}
\includegraphics[width=0.9\textwidth]{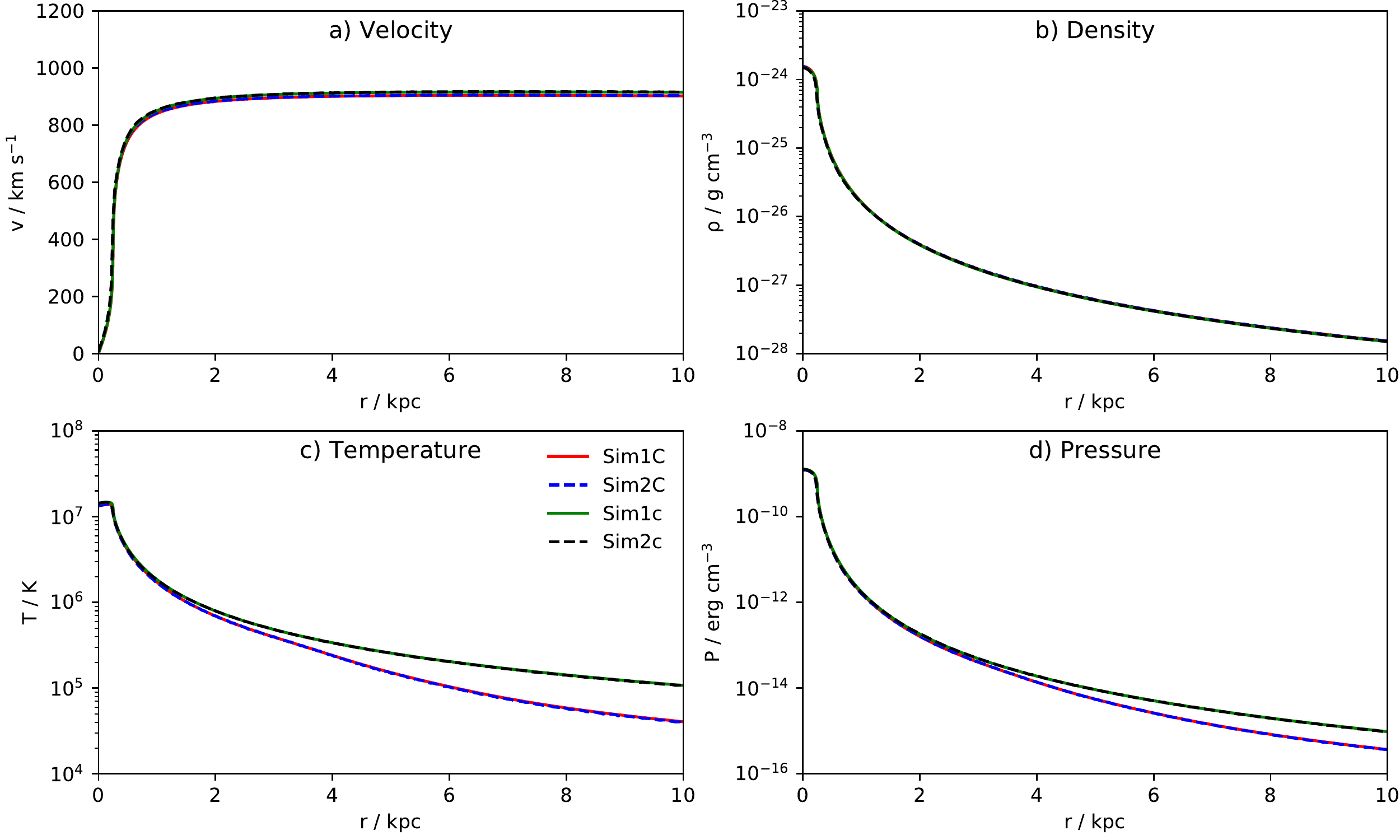}
\caption{HD profiles for comparable 1D and 2D simulations after reaching their stationary state. Outflow velocities are plotted in panel \textit{a}; gas density is plotted in panel \textit{b}; gas temperature is plotted in panel \textit{c}; gas pressures are shown in panel \textit{d}. For the 1D cases, the green lines show \texttt{Sim1c} without radiative cooling, and the red lines are the corresponding results when radiative cooling is fully taken into account (\texttt{Sim1C}). The corresponding 2D simulations are shown by the dashed black and blue lines respectively. 
}
\label{fig:HD1D_cooling}
\end{figure*}

The radiative cooling of gas in an outflow 
  generally produces one of three outcomes 
   \citep[see][]{Silich2004}.   
(1) Radiative cooling is inefficient 
     in tenuous (and hot) outflows,  
     as both free-free and bound-free processes 
     are strongly dependent on the gas density. 
  With negligible radiative energy loss, 
    the flows are indistinguishable from 
 flows where radiative cooling is not considered. 
(2) Radiative cooling is effective in warm, dense outflows. 
Severe radiative cooling 
  can cause fragmentation in the outflow, 
  and clumps are formed under self-gravity.  
This fragmentation would destabilise an outflow,  
  and those with strong radiative losses 
  are often unstable. 
Clumpy outflows are multi-phase fluids.  
The complexity in the phase transitions and interactions, 
  together with HD instabilities, 
  make the outflow highly variable. 
(3) Between the two extreme situations above 
  are cases where cooling in the flow 
  is significant, 
  but not substantial enough 
  to prevent a stationary state from being reached.
The outflow in simulation \verb|Sim1C|, 
 is an example of this scenario. 

Figure~\ref{fig:HD1D_cooling} shows the comparison 
   of outflows with cooling (\verb|Sim1C|) and without cooling 
   (\verb|Sim1c|), 
   after they have reached their stationary states.  
The stationary-state temperature and pressure profiles 
  of an outflow are modified in the presence of cooling. 
As shown, the temperature and pressure in the outflows 
  are lower in \verb|Sim1C| than in \verb|Sim1c|
  at large radii,    
  and in \verb|Sim1C| the gas temperature of the outflow 
  shows a noticeable drop 
  near $r\sim4\rm\ kpc$, at $T\sim 3\times 10^{5}~{\rm K}$. 
The velocity profile is slightly adjusted 
  when radiative cooling is present  
  (as the thermal pressure would be reduced, hence altering the pressure force,    
  when some portion of thermal energy in the gas is radiated away). 
The density is, however, less affected by radiative cooling.

\subsubsection{Comparison between 1D and 2D simulations} 
\label{sec:1D2D}

2D flows and 1D flows 
  can show very different thermal and dynamical properties.    
These differences can be manifested in their HD instability. 
For example, Kelvin-Helmholtz instabilities and Rayleigh-Taylor instabilities,  
   commonly present in astrophysical flows, 
   occur in 2D flows but are not allowed in 1D flows. 
There are certain instabilities 
  that can occur in 1D as well as 2D flows --   
  an example being the thermal instability  
  in non-relativistic shocked heated flows 
  in accreting magnetic compact objects    
  \citep[e.g.][]{Chevalier1982ApJ,Saxton1998MNRAS}. 
While dimensionality-introduced instabilities  
  can be important in time-dependent flows, 
  they would either reach saturation or be damped out 
  if the system can evolve into a stationary state. 
The galactic outflows investigated in this study 
  could evolve into a stationary state, at least locally, 
  for certain conditions appropriate for starburst galaxies. 
It is therefore useful to compare  
  the resulting HD structures 
  of the outflows obtained in 1D and 2D simulations. 
In the comparison we consider 
  the 1D simulated flows   
  (\verb|Sim1C| with radiative cooling and  \verb|Sim1c| without radiative cooling), 
  which assume a spherical symmetry, 
  and 2D simulated flows 
  (\verb|Sim2C| with radiative cooling and \verb|Sim2c| without radiative cooling), 
  which assume a cylindrical symmetry, 
  when they reach their corresponding stationary states. 
As shown in Figure~\ref{fig:HD1D_cooling}, 
  the stationary-state profiles of the 1D and 2D outflows, with or without radiative cooling,  
  agree remarkably well. 
Their agreement is of astrophysical importance,  
  as it implies that 
  1D simulations are sufficient to generate templates 
  of galactic outflow HD profiles 
  in post-processing radiative transfer calculations, 
  for the systems that reach a (quasi-)stationary state.

\begin{figure*}
\includegraphics[width=0.9\textwidth]{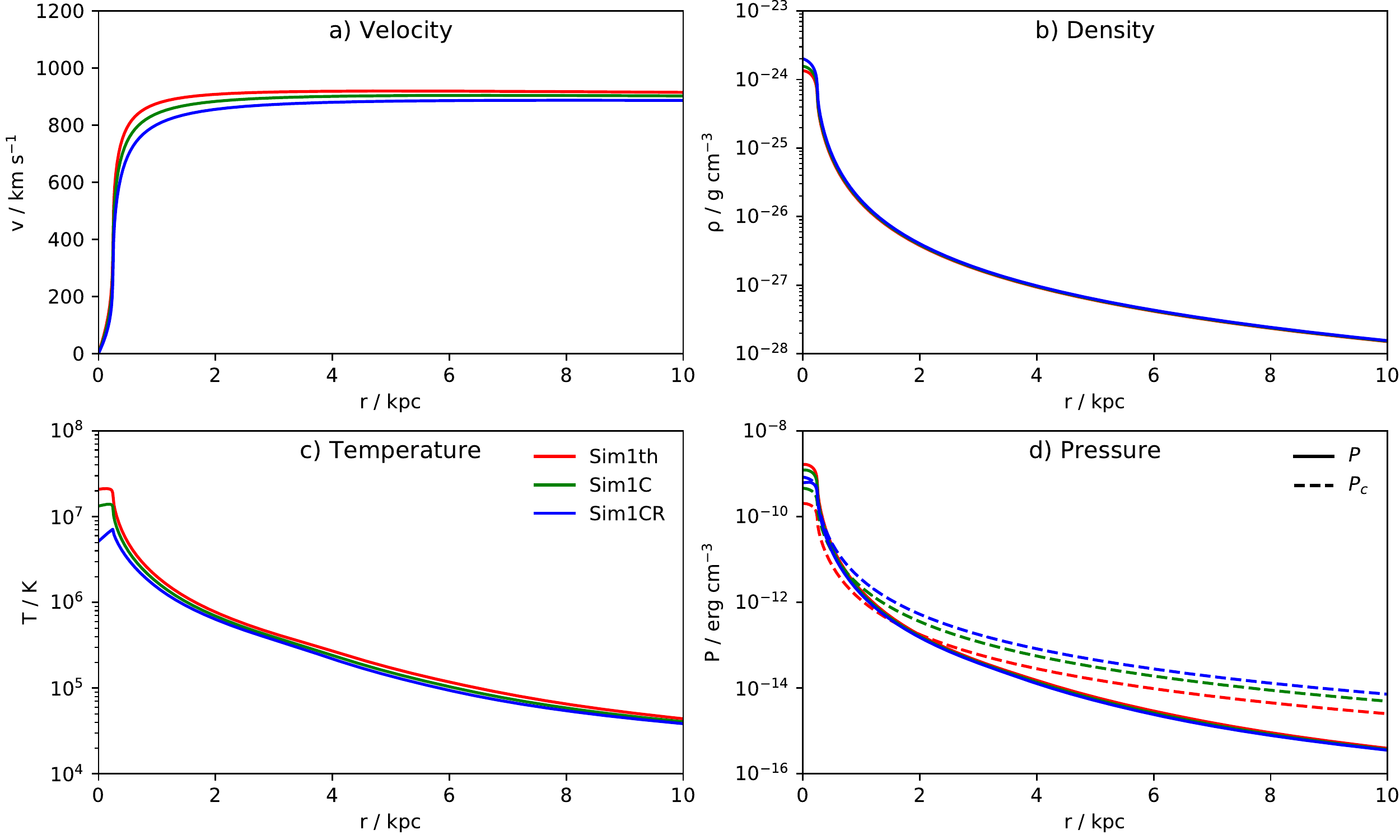}
\caption{Stationary-state HD profiles for 1D simulations driven predominantly by thermal pressure, CRs and a combined scenario including both of these. Outflow velocity, density, temperature, and pressures are plotted in panel \textit{a}, \textit{b}, \textit{c}, and \textit{d} respectively. 
The red, blue, and green lines show \texttt{Sim1th} 
  (thermal-mechanically driven), 
  \texttt{Sim1CR} (CR driven), 
  and \texttt{Sim1C} (both equal) respectively.}
\label{fig:HD1D_mechanism}
\end{figure*}

\subsection{Driving mechanisms}
\label{sec:CR}

As radiative pressure is inefficient in driving the outflow 
  in general situations,    
  this study puts focus in 
  outflows driven by thermal mechanical pressure and/or CRs.    
Table~\ref{tab:sim} lists the configuration 
  of the simulations   
  (and the HD formulation 
  can be found in \S~\ref{sec:injection_terms}).  
The stationary-state results were already presented in \citetalias{Yu2020}, 
  which showed, in general,  
  thermal mechanical pressure is an effective mechanism 
  to accelerate the gas in the outflows.   
The efficiency of CRs is weaker in comparison,   
  and the gas at the base of the outflow has lower temperatures 
  relative to the thermally-driven case. 
  This can be seen in Figure~\ref{fig:HD1D_mechanism}, where the radial HD profiles of \texttt{Sim1th} 
  (thermal-mechanically driven), \texttt{Sim1CR} 
  (CR driven), and \texttt{Sim1C} (both) 
  are shown. While the temperature of the central region in \texttt{Sim1CR} is low, the CRs can be seen to provide a driving effect at larger distances, via Alfv\'{e}nic coupling. 
As a consequence, the gas is not only accelerated but also heated. 
This results in a shallower temperature gradient in an outflow at large galactocentric distances.
\subsection{Star formation rate}  
\label{sec:SFR_evolution}

\subsubsection{Stochastic variation}
\label{sec:noise}  

In the simulations, 
  when a fixed SFR ($\mathcal{R}_{\rm SF}$)
  is used to parametrise  
  the injection of energy and mass from the starburst core 
  into the outflows,   
  a time-averaged rate 
  (associated with star formation and hence SN activity)  
  is implicitly adopted. 
This implementation does not always fully capture 
  the full picture of how galactic outflows 
  are determined by the star formation activity 
  in the galaxies, which, in reality, is not stationary.  
Time variations in the energy injection and their effects 
  are addressed in the simulations \verb|Sim1P| and \verb|Sim2P|, 
  by introducing fluctuations of $\mathcal{R}_{\rm SF}$  
  in each simulation time step.  
Stochastic variations are assigned to SN events, 
  specified by a rate $\dot{N}_{\rm SN}$, 
  assuming a Poisson process.  
The sole parameter of these Poisson fluctuations is 
  the expected value of the SN rate, 
  $\langle\dot{N}_{\rm SN}\rangle$ 
  (and this value can be considered as statistically equivalent 
  to the time-averaged value of the SN event rate).  

A SN event rate 
 $\dot{N}_{\rm SN} =0.2\rm\ yr^{-1}$ 
 \citep[appropriate of M82-like galaxies, see e.g.][]{deGrijs2001, Veilleux2005} 
 is adopted in the simulations \verb|Sim1P| and \verb|Sim2P|.
 As the simulated outflow approaches stationary state, the time step is roughly 250 yr, which corresponds to an expected value of 50 
  SN events in each time step, 
  with a variance of 50 events 
  for Poisson fluctuations.   
The HD profile of the outflows 
  obtained in \verb|Sim1P| and \verb|Sim2P| 
  are indistinguishable 
  from those of \verb|Sim1C| and \verb|Sim2C| respectively, 
  implying that 
  stochastic amplitude variations, 
  following a Poisson process,    
  on time scales shorter than 
  the dynamical time of the outflow ($\sim 10~{\rm Myr}$)  
  have no significant effects 
  on the asymptotic properties of outflows 
  that have reached a stationary state.   
Hence, 
  assuming a parametric mean SFR is sufficient  
  for the general application purpose 
  when the variations are not on timescales  
  comparable to the dynamical timescale of the outflow.

\begin{figure*}
\vspace*{0.5cm}
\includegraphics[width=\textwidth]{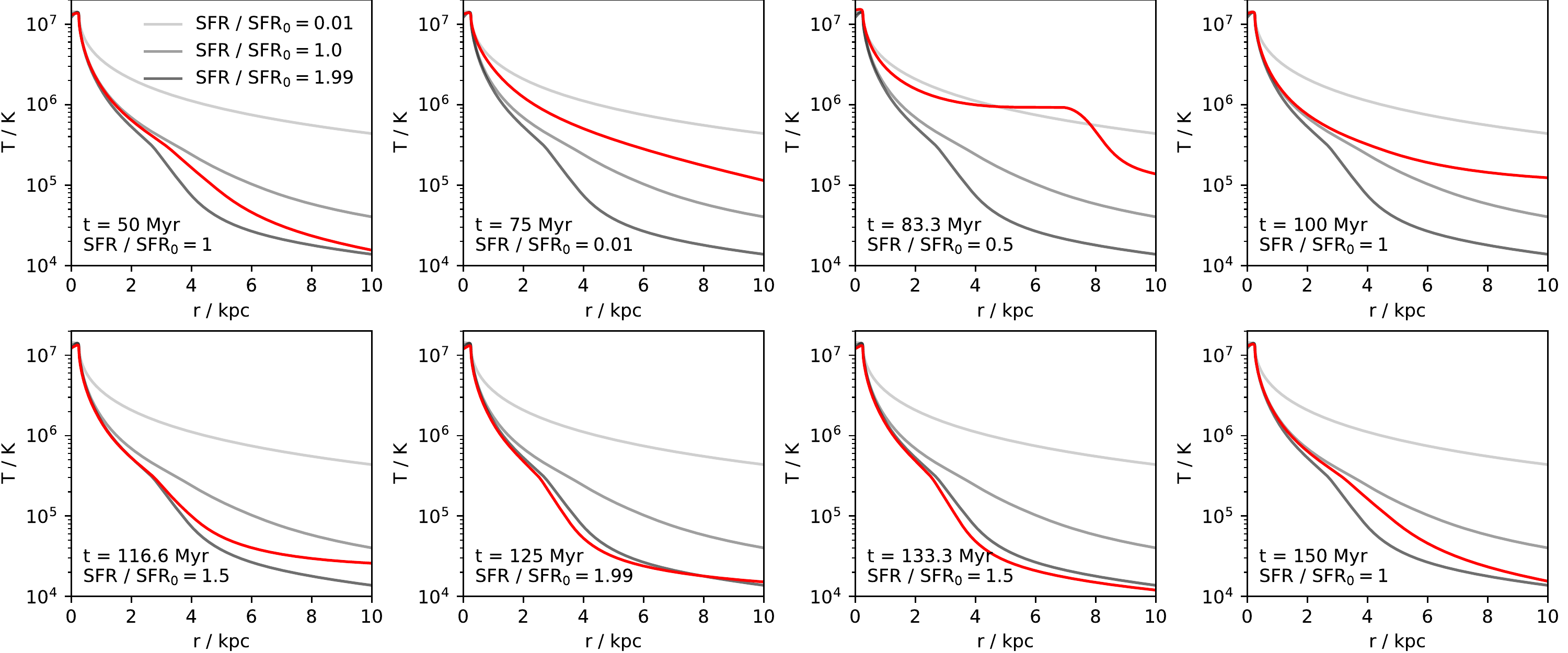}
\vspace*{-0.2cm}
\caption{Temperature profile for \texttt{Sim1S} 
(where cyclic variation in the SFR is considered) within 1 duty cycle from 50 to 150 Myr, overlaid against the analytical stationary-state solutions at $\mathcal{R}_{\rm SF}/\mathcal{R}_{\rm SF, 0}=0.01$, 1, and 1.99, representing the bottom, mean, and peak SFR of the variation cycle respectively.
Note that the temperature appears to be lower for higher SFR  
 for the parametrised approach we have adopted, i.e. $Q/q$ is kept constant. 
The flow is a adjusted accordingly to satisfy the conservation of momentum and energy 
 and the imposed boundary condition. 
The cooling rate is not a monotonic function of temperature.  
The gas temperature of the flows shown here is apparently lower when the SFR is higher, 
  because the combination of their densities and temperatures  
  happens to give more efficient radiative cooling.}
\label{fig:HD1D_sin}
\end{figure*}

\subsubsection{Episodes of starburst and quenching}
\label{sec:SFR}  

The SFR in galaxies may vary over time, 
  on timescales comparable to 
  or longer than the dynamical time of the outflows. 
The burst of star formation in M82, 
  induced by its last tidal encounter with M81 
  has substantially subsided, 
  as indicated by age analyses 
  of its the stellar population 
  \citep[see][]{deGrijs2001,Mayya2004ApJ}.   
Some proto-galaxies,   
  e.g. MACS1149$-$JD1, at $z \approx 9.1$ 
   \citep{Hashimoto2018Nat}, 
  had already exhibited multiple episodes 
  of rapid star formation and quenching   
  within a very short interval 
  of a few hundred Myr 
  \citep[see e.g.][for a possible mechanism to quench star formation in very young galaxies in the early Universe]{Owen2019b}.

The effects of episodic  
  star formation and quenching 
  on the HD properties 
  of galactic outflows 
  are investigated 
  in the simulation \verb|Sim1S|.  
Figure~\ref{fig:HD1D_sin} shows the 
  temperature profiles in outflows    
  obtained by \verb|Sim1S| 
  at eight different times.   
The simulation adopts a cyclic variation in the SFR, 
  modelled by a sinusoidal function with cycles of 100~Myr, 
  a timescale appropriate for merger-driven starbursts  
  \citep[see e.g.][]{diMatteo2008A&A}, 
  as described in equation~\ref{eq:sin}.
The variations in the temperature profiles over time,  
  though still a cyclic process, 
  have a certain subtleness  
  because of the response of the gas in the outflow  
  to the time-dependent boundary conditi on 
  (via the cyclic SFR variations). 
The first and the last panels 
  correspond to 
  the initial and last evolutionary stage 
  of a 100-Myr cycle.  
Before this cycle begins, 
  the initial shock
  has already propagated out of the simulation domain.  
The simulation shows that 
  the outflow temperature 
  at a given time and position 
  can be approximated 
  by the value obtained from an analytic calculation 
  at its designated SFR,  
  with a time delay that is proportional 
  to the galactocentric distance $r$.  
This delay is characterised by the dynamical timescale 
  of the outflow, 
  which is approximately given by 
  $t_{\rm out}=r/v_{\rm st}(\infty)$ (similar to \S~\ref{sec:general}). 
Simulations of flows with 
  shorter SFR cycles (10 and 30 Myr) were also conducted,  
but we found that the behaviours are qualitatively similar 
  to the cases presented above,  
  with the timescales of variations in the HD variables 
  shortened accordingly.

\subsection{Geometry of the starburst region}
\label{sec:shape}  

The morphology of outflows is predominantly governed 
   by the path of least resistance 
   encountered by a developing wind in the ISM of the galaxy.
This typically result in a bi-conical structure \citep{Veilleux2005}.  
So far a spherical starburst core has been considered in the simulations.  
If the starburst core deviates substantially from a sphere, 
  this path of least resistance would alter, 
  thus modifying the development of outflow 
  and hence its morphology and HD structure.  
Although 3D simulations with a realistic ISM distribution 
  would be needed in order to better capture the details of geometrical effects 
  and the convolution of geometry with astrophysical processes,   
  2D simulations can still provide useful insights into 
  how geometry of the starburst core would 
  affect the outflow that it launches  
  to engage with a surrounding ISM.  
2D simulations are therefore conducted in this study,  
  with non-spherical central star-forming regions (starburst cores).  
  
\begin{figure*}
\includegraphics[width=0.9\textwidth]{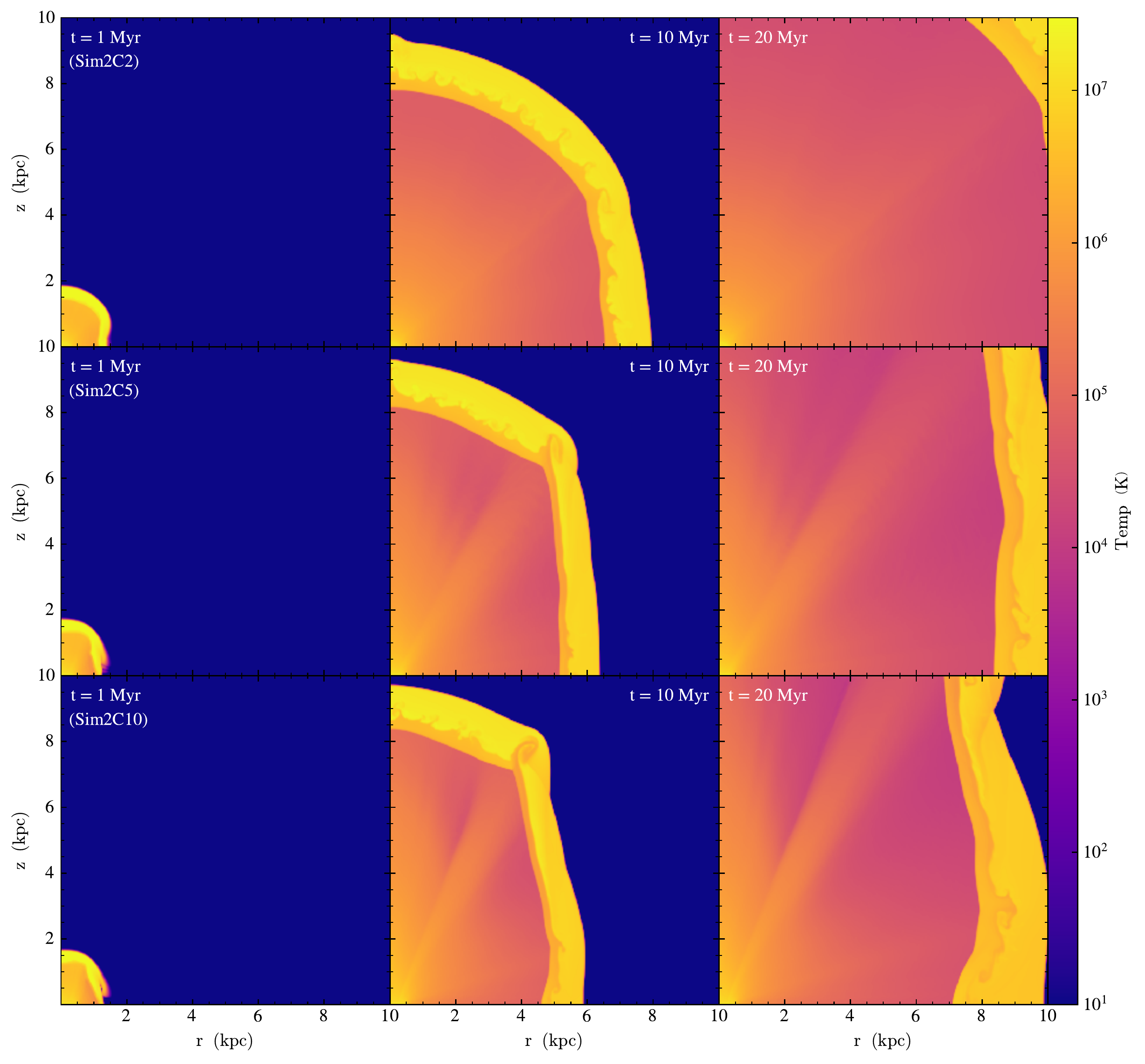}
\caption{2D temperature maps for {\tt Sim2C2}, {\tt Sim2C5} and {\tt Sim2C10}, where different shapes of the star-forming region are considered. An ellipse with aspect ratio $R_{\rm A} = 2$ is used in {\tt Sim2C2}, $R_{\rm A} = 5$ in Sim2C5, and $R_{\rm A} = 10$ in {\tt Sim2C10}. The evolution of the flow is shown at 1, 10 and 20 Myr.}
\label{fig:HD2D_ellipse}
\end{figure*}

The shape of the starburst core 
  is a geometrical boundary condition 
  in the HD formulation.    
As 2D simulations would require 
  an axi-symmetric boundary condition, 
  axi-symmetric ellipsoids 
  are the assumed geometrical shape 
  for the starburst core.   
Thus, these starburst cores 
  can be parametrised by an aspect ratio $R_{\rm A}$, 
  whose values are assigned to be 2, 5 and 10 
  for the simulations 
   \verb|Sim2C2|, \verb|Sim2C5| and \verb|Sim2C10|, 
   respectively. 
Figure~\ref{fig:HD2D_ellipse} 
  shows the 2D temperature maps of the outflows 
  in the simulations, 
  at 1 Myr, 10 Myr and 20 Myr after their onset. 
The asymmetric morphology and development of the flows 
 can be seen,   
 as the emerging of outflow echoes 
 the geometrical shape of the star-forming region.  
Though there are similarities among the three simulated outflows, 
  the connection between core shape and outflow shape 
  is not particularly strong 
  (cf. \verb|Sim2C5| and \verb|Sim2C10|, 
  whose values of $R_{\rm A}$ differ by a factor of 2). 
This indicates that 
  the shape of the star-forming region 
  (the geometrical boundary condition) 
  is not a major factor 
  to determine the outflow morphology and development. 
The path of least resistance   
  is determined mainly 
  by the conditions exterior to the star-forming core, 
  i.e. the physical conditions in the ambient ISM 
  and the global energy injection into the gas  
  and redistribution within the outflow. 

\subsection{Comparison with observations}
\label{sec:m82_compare}

Simulations in this work have adopted 
  parameters appropriate for M82-like galaxies. 
We therefore consider M82 as a reference 
  for comparison with observations. 
M82 was observed in X-rays  
  on many occasions \citep[e.g.][]{Ranalli2008,Konami2011,Lopez2020}.  
Outflows from starburst galaxies, e.g. NGC~253 and M82   
  generally have several ionised gas components 
  with different thermal temperatures, 
  which is evident in the multi-temperature fits 
  to their X-ray spectra.   
In M82, the presence of charge exchange lines 
  \citep{Konami2011,Zhang2014ApJ} 
  in the X-ray spectra 
  implies that the outflow in M82 must be multi-phase, 
   with also substantial amount of neutral material.

The treatment of flows with multiple thermal components 
   and phases  
   involves non-trivial processes, 
   e.g. the interplay between heating and mass exchange, 
   phase mixing, phase transitions, and the relative drag 
   between the components, 
   which are beyond the scope of this study. 
Nonetheless, some meaningful comparisons  
  can still be drawn between this work and observations, 
  under certain simplifications and appropriate approximations. 
In a broad perspective, 
  an outflow can be considered to consist 
  of a hot ionised component and a cold component. 
The former is the highly ionised 
  and generally satisfies a fluid description; 
  the latter is a mix of minimally-ionised and neutral gas 
  and clumps, 
  for which the standard HD description needs to be modified. 
As such, we restrict the comparison 
  of our results to observations of the hot component.  
  
The temperature of the hot phase may reach $\sim 10^7{\rm K}$, 
  and the observation of \cite{Lopez2020} 
  showed a temperature of $\sim3\times10^7{\rm\ K}$ 
  near the base of the outflow in M82. 
Our simulations show 
  that these temperatures can be achieved  
  if the flows are thermally driven,  
  (see simulation {\tt Sim1th}). 
They are, however, substantially higher than 
  the temperatures of the outflows driven by CRs 
  (see simulation {\tt Sim1CR}).  
The observations of \cite{Lopez2020} 
  also showed outflow temperatures in M82 
  decrease to $\sim7\times10^6{\rm\ K}$ 
  and $\sim4.5\times10^6{\rm\ K}$ 
  at $r\sim1\rm\ kpc$ and $r\sim2.5\rm\ kpc$ respectively. 
These temperatures are higher 
  than the values obtained at corresponding locations in our simulations, 
  even when radiative cooling is not considered 
  (cf. Figure~\ref{fig:HD1D_cooling}).  
As has been shown previously in \citetalias{Yu2020}, 
  the temperature of the outflow gas 
  would be altered by the input values 
  for the magnetic field strength or mass loading.  
It is possible that M82 has a lower mass load 
  and/or a more highly structured and turbulent magnetic field 
  \citep[see][]{Lopez-Rodriguez2021ApJ}
  than considered in our simulations.  
The M82 outflow is at least a two-component fluid, as indicated by 
  the observation of \cite{Lopez2020} and also others,  
  and the mass loading into the hot component 
  is expected to be lower than into the wind overall at lower altitudes, 
  before substantial evaporation of a colder, denser gas phase and mixing of the two components could arise\footnote{This scenario would be consistent with the findings of \cite{Lopez2020}, which also suggest that the hot winds are being mass loaded due to the mixing and heating of cooler entrained gas into the hot phase as it is advected with the flow.}.  
This work has put a focus on the global HD for different driving mechanisms
  and the consequential X-ray properties.  
  Complexities of multi-component, multi-phase flows 
  have not been fully accounted for.  
It is therefore not surprising that 
  there are some temperature discrepancies between our work and observations
  at large distances from the starburst core. 
With these factors taken into consideration, 
  the results obtained here 
  are not inconsistent with the X-ray observations of M82  
  and other starburst galaxies, e.g. NGC 253~\citep[e.g.][]{Bauer2007AA, Mitsuishi2013PASJ}. 

\section{X-ray emission}
\label{sec:XR_emission}

\subsection{Synthesis of X-ray spectra}
\label{sec:XR_synthesis}

Starburst galaxies are known to be X-ray sources, with X-rays being emitted from their hot ISM,
 stars, accreting compact objects (in particular, X-ray binaries), 
 SNe, 
 and galactic wind outflows 
 \citep[see e.g.][]{Fabbiano1988,Soria2002A&A,Mitsuishi2013PASJ,Lopez2020}. 
For nearby galaxies, e.g. NGC~253 and M82,  
  where outflows can be resolved 
  in X-ray observations,  
  there have been attempts to infer the outflow HD profiles
  from X-ray imaging data.  
\cite{Lopez2020} derived the temperature profiles 
  of the outflows in M82    
  from the variation of their X-ray surface brightness  
  across galacto-centric distances along their minor axis, 
  and the X-ray profiles of galactic outflows, 
    with mass loading, were also computed very recently 
    by \cite{Nguyen2021arXiv} using numerical HD models.

Outflows in distant starburst galaxies 
  are, however, not always resolvable spatially 
 in X-ray imaging observations. 
Despite this, 
  spectral information in X-rays from outflows 
  observed in such distant galaxies would still be retained 
  and, hence, the thermal properties of the gas 
  in their outflows could be inferred from this. 
In particular, 
  the relative fluxes in different, properly selected X-ray bands 
  can provide constraints for the physical conditions 
  within X-ray emitting outflow gas. 
With this in mind, 
  we compute synthetic X-ray spectra 
  for the simulated galactic outflows considered in the previous sections.   
Solar metallicity is adopted in the X-ray spectral calculations, 
  unless otherwise stated.
The procedures for computing X-ray spectra 
  in outflows from star-forming protogalaxies 
  and young galaxies at high redshifts  
  are the same as those adopted for nearby outflows, 
  with only the metallicity being modified to appropriate levels. 

X-ray spectra are computed using \verb|APEC| (version 3),   
  with the density and temperature inputs  
  provided by the HD simulations\footnote{The emission volume 
    of the outflowing gas is   
$V= \Delta \Omega  \int_{r_1}^{r_2} {\rm d}r ~(r^2)$,      
  where $\Delta \Omega$ is the solid opening angle of the outflow cone. 
In spectral computations, 
 we set the inner boundary $r_1 = 1$~kpc  
 and the outer boundary $r_2 = 10$~kpc 
 (\S~\ref{sec:XR_cooling_evolution}).  
For optically thin emission, 
  the spectral properties are independent of $\Delta \Omega$.  
Hence, we simply set $\Delta \Omega = 4\pi$ without losing generality. 
The total luminosity of the outflow should therefore 
  be scaled with a factor equal to the 
  ``real'' solid opening angle of the outflow cone divided by $4\pi$ radians.}, 
  with energies $E$ between 0.1 and 10~keV 
  and resolution of 0.01~dex 
 (i.e. a $\log_{10}(E)$ resolution of 0.01).
The chosen spectral energy range is appropriate
  for thermal emission 
  associated with bound-free and free-free  
  processes from optically thin gas 
  at temperatures $T \sim 10^5-10^7~{\rm K}$, 
  as indicated 
  by observations of the hot ISM  
  in nearby starburst galaxies, e.g.    
  NGC~253 \citep{Mitsuishi2013PASJ} 
  and M83 \citep{Soria2002A&A}. 
It is also the spectral energy range 
  where the X-ray observatories {\it XMM-Newton} 
  and {\it ATHENA} 
  would have large effective photon collecting areas 
  and good spectral resolving capability 
  \citep[see][]{Jansen2001A&A,Barcons2017,Barret2020AN}.  
  
For broadband spectral calculations, 
  four energy bands, (0.1$-$0.5)~keV, (0.5$-$1.0)~keV, 
  (1.0$-$2.0)~keV and (2.0$-$10.0)~keV are selected.  
These bands correspond respectively to spectral regions  
  associated with Galactic absorption, 
  extra-galactic line-of-sight absorption, 
  thermal bound-bound emission from optically thin hot gas 
  at $T \sim 10^5 - 10^7{\rm K}$,  
  and emission from stellar X-ray sources 
  that can contaminate the emission from the outflow. 
In each energy band, 
  the ''luminosity'' is obtained 
  by summing the emission 
  over the energies within the band. 
  
\begin{figure}
\vspace*{0.1cm}
\includegraphics[width=\columnwidth]{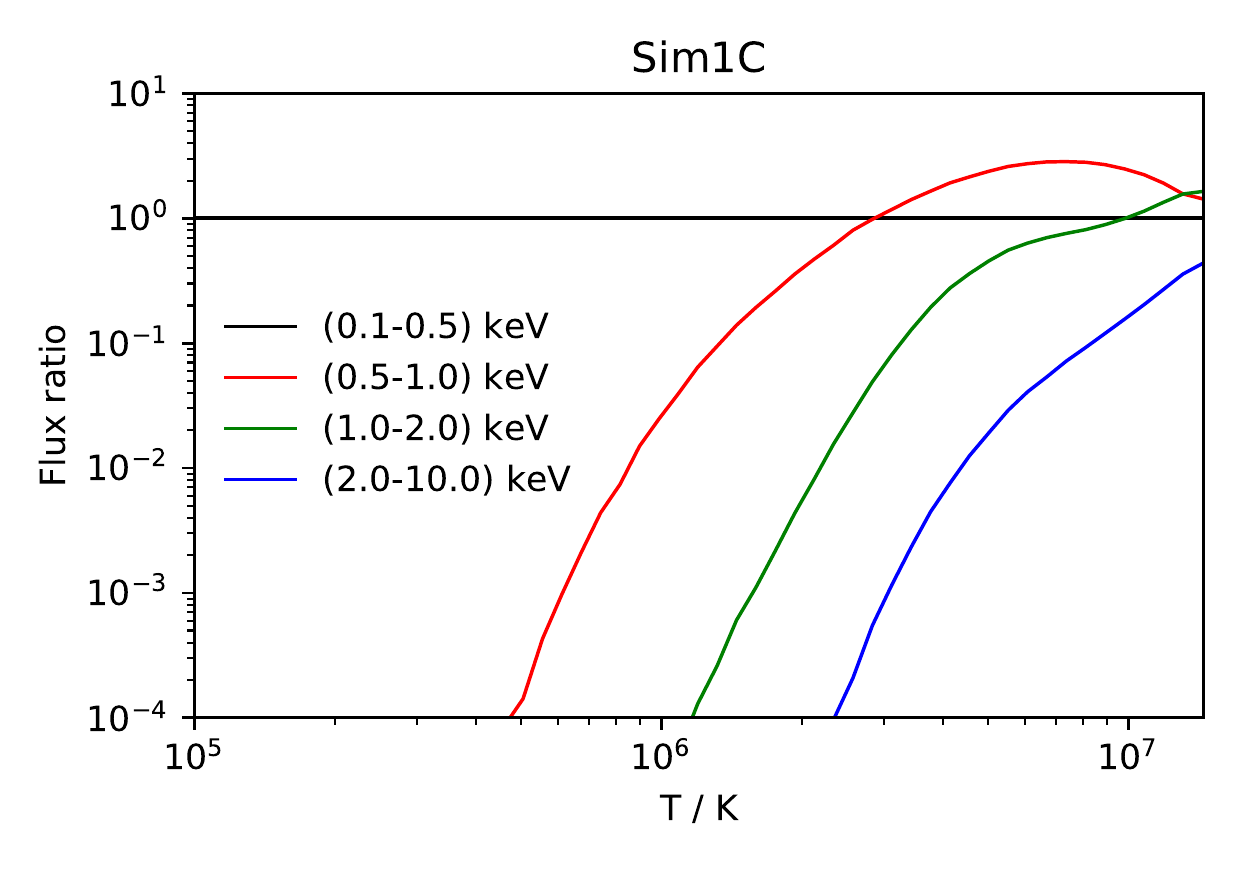} 
\vspace*{-0.5cm}
\caption{Contribution of outflow gas between temperatures of $10^{4}-10^{7} ~{\rm K}$ to the four X-ray bands, calculated with {\tt Sim1C} at 20 Myr (i.e. after the stationary-state configuration has been attained). Most of the emission would be associated with the lower $\sim$ 4 kpc of the flow.
 The relative contribution is expressed in terms of a flux ratio 
  between the flux in the specific energy band to the flux 
  in the $(0.1-0.5)$~keV band.} 
\label{fig:dLxdT_regions}
\end{figure}

Galactic outflows are composite plasmas  
 with a wide spread of temperatures.   
The X-ray luminosities of an outflow 
 in the 4 energy bands specified above 
 contain emission from the entire flow, 
 although more contribution comes from regions 
 where the gas temperatures match  
 the thermal temperature for 
 the emission flux to peak within the energy range of the band. 
Figure~\ref{fig:dLxdT_regions} illustrates  
 the relative contribution of gas with different temperatures 
 to the energy bands. 
As expected, 
  cool gas (with $T < 10^6{\rm K}$)  
  dominates the emission in the ($0.1-0.5$) keV band, 
  while only the hot gas (with $T \sim 10^7{\rm K}$ and above)  
  would make a significant contribution to the $(2-10)$ keV band.  
Overall, most of the X-ray emission originates 
  from gas at temperatures 
  $\sim (5\times 10^5 - 5\times 10^6)~{\rm K}$,   
  which is associated with the inner $\sim$ 4 kpc of the outflow 
(see Figure~\ref{fig:HD1D_cooling})\footnote{We 
  show the outflow in simulation {\tt Sim1C} 
  in Figure~\ref{fig:dLxdT_regions} as an illustration. 
The outflows of the other simulations considered in this work 
  produce qualitatively similar temperature-emission curves. 
For instance, 
  {\tt Sim1th} yields slightly lower X-ray emission 
  in the softer energy bands 
  for higher gas temperatures, above $\sim 10^6 ~{\rm K}$, 
  and slightly higher X-ray emission in the higher energy bands 
  than {\tt Sim1C}.}.  
  
\subsection{X-ray spectroscopy}
\label{sec:XR_spectroscopy}

\subsubsection{Cooling and outflow development}
\label{sec:XR_cooling_evolution}

\begin{figure}
\vspace*{0.1cm}
\includegraphics[width=\columnwidth]{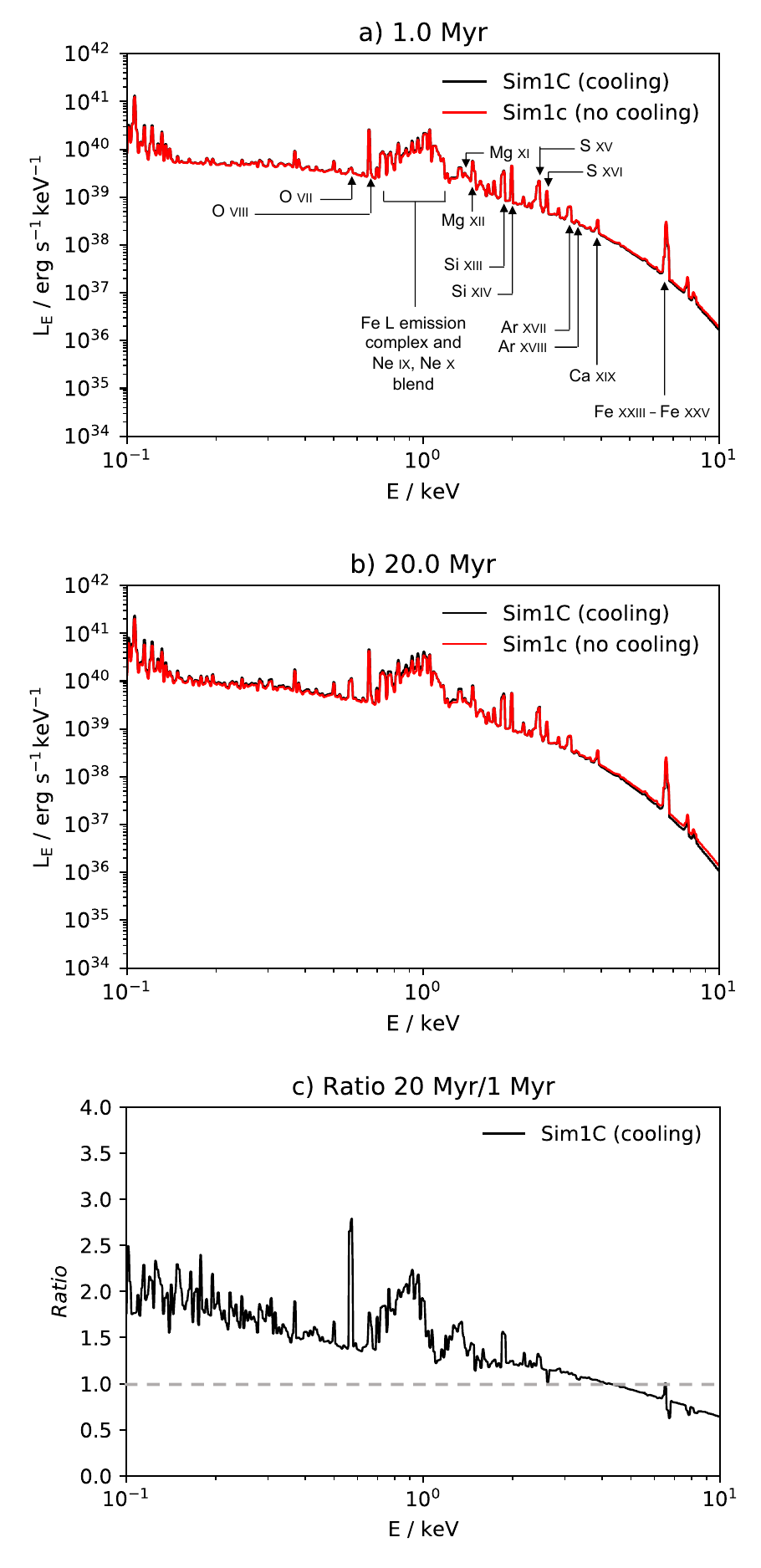} 
\vspace*{-0.5cm}
\caption{Synthetic spectra, at energies (0.1$-$10)~keV,  
  of the X-ray emission from outflows 
  in simulations {\tt Sim1c}  
  (which ignores radiative cooling) 
  and {\tt Sim1C} (which includes radiative cooling).   
The spectra at the top and in the middle panels   
  (panel a and panel b respectively)   
  are the X-ray spectra at 1~Myr and 20~Myr   
  after the outflows are launched from the starburst core. 
The spectrum in the bottom (panel c)  
  is the quotient spectrum at 20~Myr with respect to that at 1~Myr   
  for the outflow in the simulation {\tt Sim1C}. The K$\alpha$ emission lines 
  commonly present in astrophysical coronal plasmas 
  at keV temperatures 
  are annotated with the spectra in top panel.}
\label{fig:xray_sim1cC_compare}
\end{figure}

Figure~\ref{fig:xray_sim1cC_compare}, 
   panels (a) and (b), 
  shows the computed X-ray spectra  
  in the $(0.1-10)~{\rm keV}$ energy range,     
  from the outflows 
  in simulations \verb|Sim1c| (which has no cooling) 
  and \verb|Sim1C| (with cooling), 
  at two epochs, 
  1~Myr and 20~Myr after the launch of the outflow.   
In both cases,  
  the emission region is restricted 
  to the main body of outflow, 
  which spans from  $1~{\rm kpc}$ to $10~{\rm kpc}$, 
  measured radially from 
  the center of the galaxy.  
The emitting gas has a solar abundance 
  \citep[cf. the metallicity in M82,][]{Origlia2004ApJ}. 
A quotient spectrum of the outflow in simulation \verb|Sim1C| 
  is also shown in Figure~\ref{fig:xray_sim1cC_compare}, panel (c),  
  for comparison of the X-ray emission 
  at 20~Myr to that at 1~Myr.

\begin{figure}
\includegraphics[width=\columnwidth]{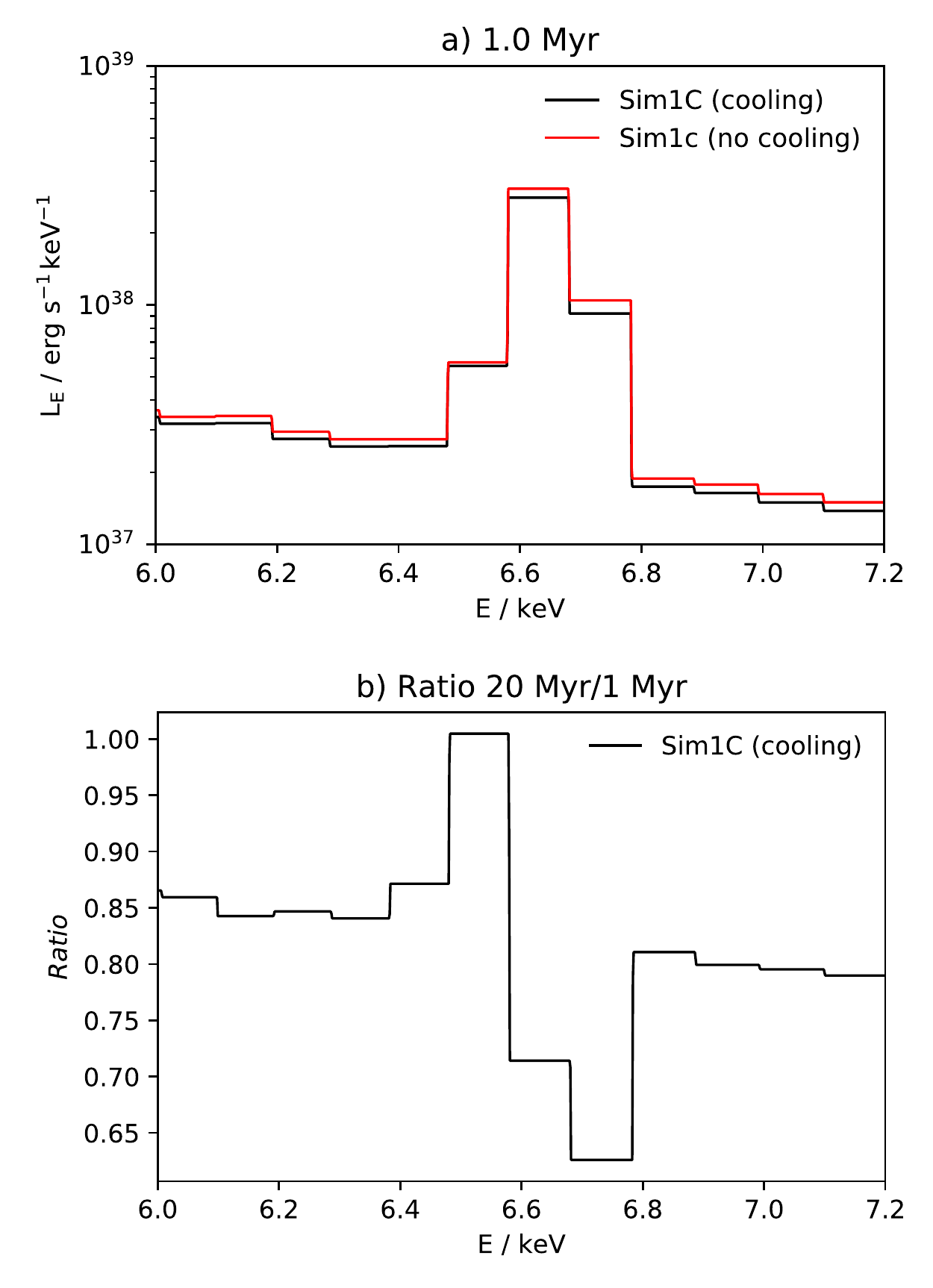} 
\vspace*{-0.5cm}
\caption{The Fe K$\alpha$ line   
  of the outflows in simulations {\tt Sim1C} 
  and {\tt Sim1c} at 1~Myr (top panel) 
  and the quotient spectrum (bottom panel) 
  at energies between 6.0 and 7.2~keV 
  for the outflow in {\tt Sim1C} 
  at 20~Myr with respect to 1~Myr. } 
\label{fig:profiles_compare}
\end{figure}

The X-ray spectra of the simulated outflows 
  are typical of 
  that of 
  a multiple-temperature coronal plasma  
  \citep[see e.g.][for X-ray emission and the associated electronic transition processes in coronal plasma]{Mewe1985A&AS,Kaastra1993A&AS,Bryans2009ApJ,Porquet2010SSRv}. 
Overall, the spectra of the outflows 
  in simulations \verb|Sim1c| and \verb|Sim1C| 
  are similar. 
The most noticeable spectral feature 
  is the broad bump at energies slightly below 1~keV.     
This is the emission complex 
  arising from the L-shell transitions of Fe\footnote{The 
  Ne K$\alpha$ emission lines are blended with the 
  Fe L emission complex.}\citep[see e.g.][]{Bryans2009ApJ,Foster2012ApJ}.   
Another striking feature is a sequence of 
  H-like and He-like K$\alpha$ emission lines 
  of O, Mg, Si, S, Ar, Ca and Fe ions.  
The Fe K$\alpha$ emission line 
  is located at around $6.6 - 6.7$~keV 
  (Figure~\ref{fig:profiles_compare}, top panel).   
There is insignificant line emission at 6.4~keV,   
  which corresponds to emission from neutral Fe  
  and weakly ionised Fe species. 
The 6.95~keV line 
  corresponding to emission from H-like Fe ions 
  is absent. 
The Fe K$\alpha$ emission  
 is therefore contributed mostly by He-like 
 (which has a closed-shell configuration in the ground state) 
 and Li-like Fe ions,    
 some from Be-like Fe ions  
 and a small amount from B-like Fe ions.   
There is a relative excess 
  in the Fe L emission complex, 
  and a slightly weaker continuum 
  at higher energies (above 4~keV) 
  in the outflow in simulation  \verb|Sim1C| 
  when compared to that in simulation \verb|Sim1c|.  
This result is not too surprising,    
  as lowering the temperature of the gas 
  by radiative cooling 
  reduces free-free emission 
  but enhances L-shell transitions.

The evolution of the X-ray spectrum  
  as an outflow progresses 
  is not obvious through visual comparison.    
The differences between the spectra 
  are better distinguished using quotient spectra,  
  which are the ratios between a spectra at a specifically chosen evolution stage,
  and a reference spectrum.  
At 1~Myr there is a shock in the outflow, 
  located about 2~kpc from the starburst core. 
By 20~Myr, the shock has already propagated beyond 10~kpc.

The shock is the outer boundary of the body 
  of X-ray emission gas in the outflow, 
  and the total power in the X-rays 
  is determined by the emission volume\footnote{For bound-bound 
  and free-free processes 
   in a collisional ionised plasma, the power is determined by the emission 
   measure, which is proportional to $\rho^2 V$, where the volume of the emission region $V$  
   would have a $r^3$ proportionality. 
   Mass continuity requires that $\rho v(4\pi r^2) ={\rm constant}$. Thus, $\rho^2 \propto r^{-4}$ and the emission is dominated 
   by the high density gas at the base of the outflow, 
   if the emission from the high-temperature 
   compressed shock-heated gas is not present.}, 
   which depends on the shock location in the transient stage.  
The difference in the emission volumes at 1~Myr and 20~Myr 
  is reflected in the ``specific intensity'' in the quotient spectrum,  
  in Figure~\ref{fig:xray_sim1cC_compare}.  
This shows values exceeding unity  
  in most part of the quotient spectrum, 
  except beyond about 4~keV.  
  The value of the ``specific intensity'' falls below unity at these higher energies because of the absence of the hot gas associated
  with the shock transition layer 
  in the outflow at 20~Myr, despite some compensation by the increase 
  in the total emission volume.
   
The most obvious feature in the quotient spectrum 
  is the Fe L emission bump at around 1~keV.  
This indicates that the Fe L-shell transitions 
  become more prominent when the outflow approaches its stationary state.    
Except for the He-like Si line, 
  almost the entire sequence of K$\alpha$ emission 
  of Mg, Si, S, Ar and Ca (all of which are clearly present 
  in the spectra at all evolutionary stages)   
  has become almost unnoticeable in the quotient spectrum, 
  indicating that the strength of the K$\alpha$ emission 
  from these species is essentially unchanged. 
There is, however, a stronger presence of the emission from  
  He-like \ion{O}{vii} 
  and He-like \ion{Si}{xiii} ions,
  and their asymmetric line profiles  
  suggests that these lines are complex, consisting 
  of z forbidden, (x,y) inter-combination 
  and w resonance components 
  \citep[see e.g.][]{Silver2000ApJ}.   
Interestingly, the Fe K$\alpha$ emission 
   in the quotient spectrum 
   has a P Cyg profile. 
However, 
  the depression in higher energy wing of the F$\alpha$ emission 
  is not caused by line-of-sight absorption. Instead, 
this is a decrease in the strength of 
  He-like, Li-like and Be-like Fe emission 
  accompanied by an increase in the strength 
  of B-like Fe emission 
  (see Figure~\ref{fig:profiles_compare}, bottom panel).

\begin{figure}
\includegraphics[width=\columnwidth]{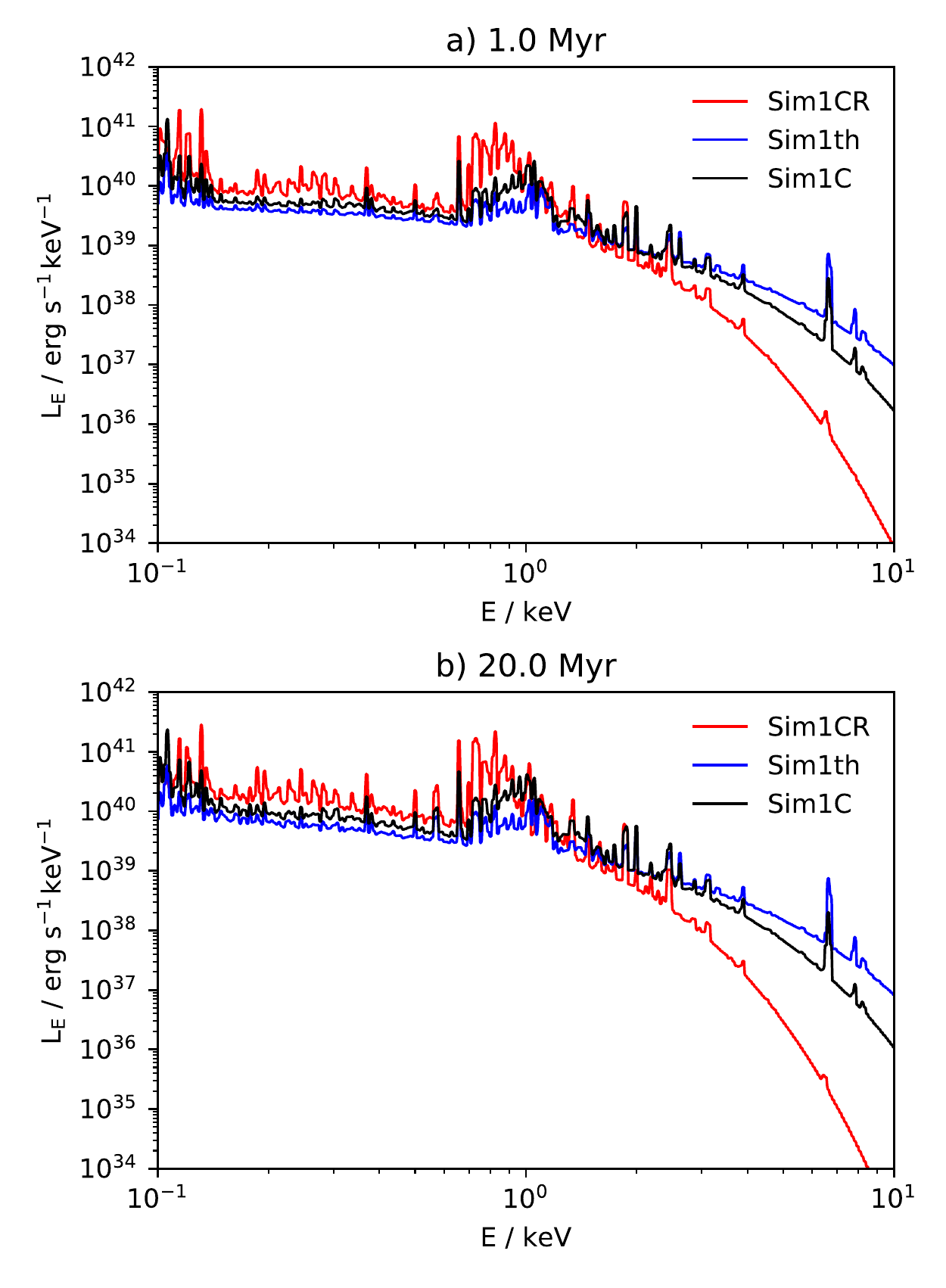} 
\vspace*{-0.5cm}
\caption{Synthetic spectra 
of the X-ray emission from the outflows 
  in simulations {\tt Sim1th} (thermal-mechanical driven), {\tt Sim1CR} (CR driven) and 
  {\tt Sim1C} (50-50 split between thermal-mechanically driven a
  and CR driven). 
Spectra in panel a correspond to 
  outflows at 1~Myr, 
  and spectra in panel b to outflows at 20~Myr.  } 
\label{fig:mechanism_compare}
\end{figure}

\subsubsection{Driving mechanisms} 
\label{sec:XR_driving}

Figure~\ref{fig:mechanism_compare} shows the X-ray spectra of outflows 
  driven thermal-mechanically (\verb|Sim1th|), by CRs (\verb|Sim1CR|) 
  and with a 50-50 split between thermal-mechanical and CR driving (\verb|Sim1C|), 
  at  
  1~Myr and 20~Myr 
  after the launch of the outflow (in panels a and b respectively).  
The Fe L emission complex 
 and the sequence of H-like and He-like 
 K$\alpha$ emission lines of 
 O, Mg, Si, S, Ar, Ca and Fe ions 
 are present in all the spectra. 
The spectra of the CR driven outflow  
 has the weakest keV emission and steepest continuum (\verb|Sim1CR|) 
  at both 1~Myr and 20~Myr,   
  while the corresponding spectrum  
  of the thermal-mechanically driven outflow (\verb|Sim1th|) 
  has the strongest keV emission and flattest continuum. 
The keV continuum weakening and steepening 
  among the outflows 
  is accompanied by a reduction in the strength of the Fe K$\alpha$ line, 
  and an enhancement of the Fe L emission complex.  
In addition, this is also accompanied by a skew of the Fe L complex towards lower energies. 
A careful inspection reveals 
  that the He-like K$\alpha$ emission, 
  in particular that of the O ions, 
  is stronger for the CR driven outflow 
  than for the thermal-mechanically driven outflow.   
Moreover, the N K$\alpha$ emission lines become more visible in the CR driven case, 
  while they are much weaker 
  in the thermal-mechanical driven outflow.  

Comparing the outflows at 20~Myr and 1~Myr 
  shows the weakening and steepening of the keV continuum emission, 
  and the accompanying change in the emission line morphology. 
This is more noticeable in the CR driven outflow 
  than the thermal-mechanically driven case. 
The overall spectral properties 
  of the X-ray emission from the the hybrid outflow (\verb|Sim1C|) 
  are between those of the CR driven and the thermal-mechanically driven cases. 
These outflow spectral variations are mostly 
  caused by the differences in the relative outflow temperatures
  (the densities of the outflows are almost identical 
  when reaching their stationary state, see Figure~\ref{fig:HD1D_mechanism}), 
  as the drop in gas temperature 
  reduces the efficiency of the free-free process,  
  but increases stronger bound-bound processes.  
Similar to the time development of the spectra of the outflows 
  in simulations \verb|Sim1c| and \verb|Sim1C| 
  (see \S~\ref{sec:XR_cooling_evolution}), 
  the factor determining the spectral evolution of each individual outflow 
  in simulation \verb|SimCR| and \verb|Sim1th|
  is the change in the temperature of the emission region,  
  which contributes the most to the total emission measure\footnote{The emission measure is the product of the square of the density 
  and the effective emitting volume.} of the outflow.

\subsubsection{Metallicity and redshift evolution} 
\label{sec:XR_redshift}

\begin{figure*}
\includegraphics[width=0.9\textwidth]{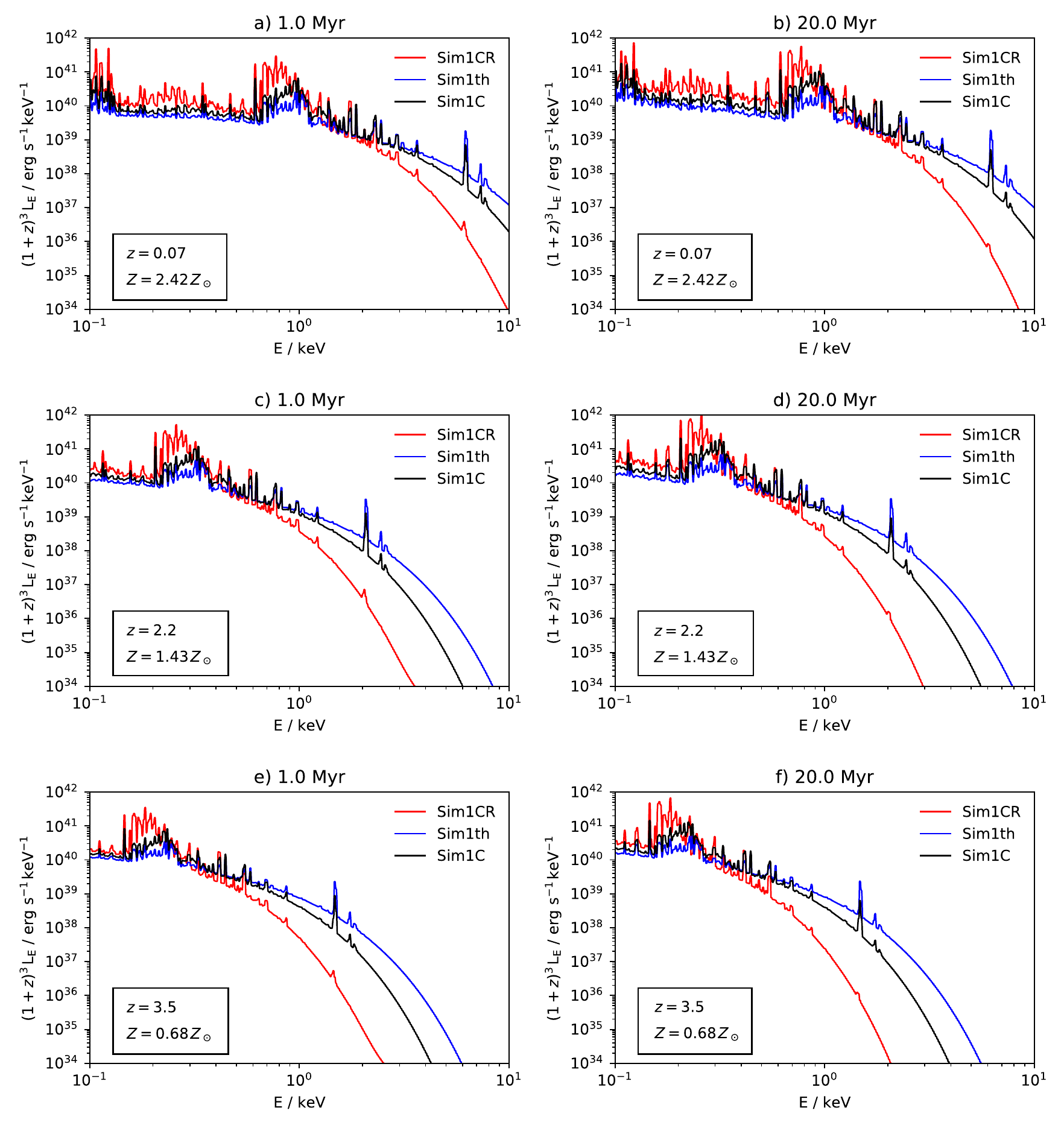}
\caption{The redshift-corrected spectra 
  of galactic outflows 
  in simulations {\tt Sim1th} (thermal-mechanically driven), 
  {\tt Sim1CR} (CR driven)  
  and {\tt Sim1C} (50-50 split between thermal-mechanical 
  and CR driven). 
Panels in the rows 
  from top to bottom show the X-ray spectra 
  at $z=0.07$, 2.2 and 3.5 respectively;   
  panels in the left and right columns show 
  the spectra of the outflows at 1~Myr and 20~Myr respectively. 
The metallicities are 
  2.42, 1.43 and 0.68 times solar abundance in the three redshift epochs, as labelled. 
  }
\label{fig:redshift_compare}
\end{figure*}
 
Features in the X-ray spectra of outflows 
  are markers of the abundances of elements, 
  although the strengths of such features are also determined 
  by the HD and thermal properties of the emitting gas. 
Abundances of elements change with the evolutionary history of the galaxies 
  and, hence, the spectra of outflows from galaxies would reflect this: 
both the emission and radiative cooling 
  are dependent on metallicity, and thus the X-ray spectra of outflows 
  would show variations over redshift     
  regardless of the underlying outflow driving mechanisms.  
This is indeed seen in the redshift-corrected spectra 
  in Figure~\ref{fig:redshift_compare}, 
  where cases are shown 
  with metallicities of 2.42, 1.43 and 0.68 times solar abundance, appropriate for their respective selected redshifts of $z= 0.07$, 2.2 and 3.5 
  \citep[see][]{Sommariva2012A&A}\footnote{These values
  are derived from scaling of the solar abundances. 
  In reality, the chemical abundances in galactic outflows are determined 
  by evolution and stellar populations in the galaxy.  
  Core collapse SN, which track star formation, 
  preferentially inject light elements (e.g. O and Ne)  
  and elements heavier than Zn~\citep[e.g.][]{Nakamura1999ApJ} 
  into the ISM and hence the outflowing gas.  
Type 1a SNe, which emerge much later,  
  inject Fe and heavy elements~\citep[e.g.][]{Nomoto1984ApJ}. 
Young galaxies at high redshifts are expected to have 
  weaker Fe L emission, but less significant reduction in strengths of 
  O K$\alpha$ lines than as in the spectra present in this work.}. 
At higher redshifts, metal abundances may be lower, leading to a decrease in 
   the overall strengths of the sequence of the 
  O, Mg, Si, S, Ar and Ca K$\alpha$ lines. 
This strength reduction is particularly obvious 
  for the two O H-like and He-like K$\alpha$ lines. 
The strength of the Fe L emission complex is also reduced, 
  but this reduction is less evident from visual inspection.
  
The keV continuum, which is mainly due to free-free processes, 
  is determined predominantly by the density  
  (and, to a lesser extent, by the temperature) of an outflow. 
\verb|Sim1c| and \verb|Sim1C| 
  show that the HD properties of outflows 
  are not particularly sensitive to variations in radiative cooling 
  (Figure~\ref{fig:xray_sim1cC_compare}; 
  see also Figure~\ref{fig:HD1D_cooling}). 
The amount of variation in the metallicity considered here 
  would not substantially alter free-free processes, 
  and hence the continuum does not show strong variation 
  between the outflows at these three redshifts. 
Despite changes in metallicity, 
  the overall trend of spectral variations among the driving mechanisms 
  is similar to that seen when assuming a solar metallicity, 
  cf. Figure~\ref{fig:redshift_compare} 
  and Figure~\ref{fig:mechanism_compare}.

\subsection{Broadband X-ray emission}
\label{sec:XR_broadband}  

\begin{figure}
\includegraphics[width=\columnwidth]{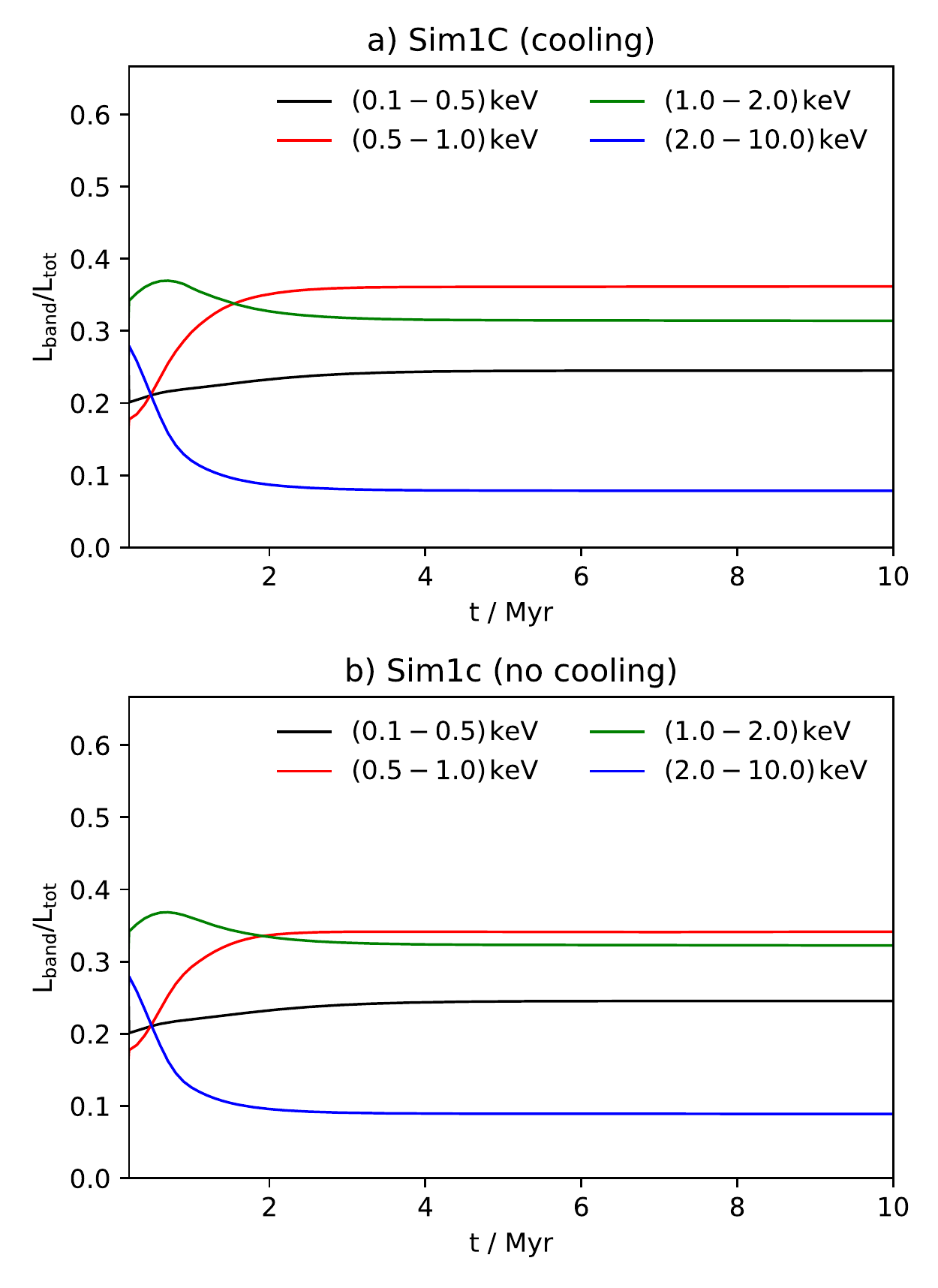} 
\vspace*{-0.25cm}
\caption{Normalised X-ray light curves of the galactic outflows 
  in simulations {\tt Sim1c} (which ignores radiative cooling) 
  and {\tt Sim1C} (which includes radiative cooling) 
  in four energy bands: $(0.1-0.5)$~keV, $(0.5-1.0)$~keV, 
  $(1.0-2.0)$~keV and $(2.0-10.0)$~keV.  
The outflows are driven by a 50-50 split between 
 between thermal-mechanical pressure and CR. 
  } 
\label{fig:sim1c_broadband}
\end{figure}

\begin{figure}
\includegraphics[width=\columnwidth]{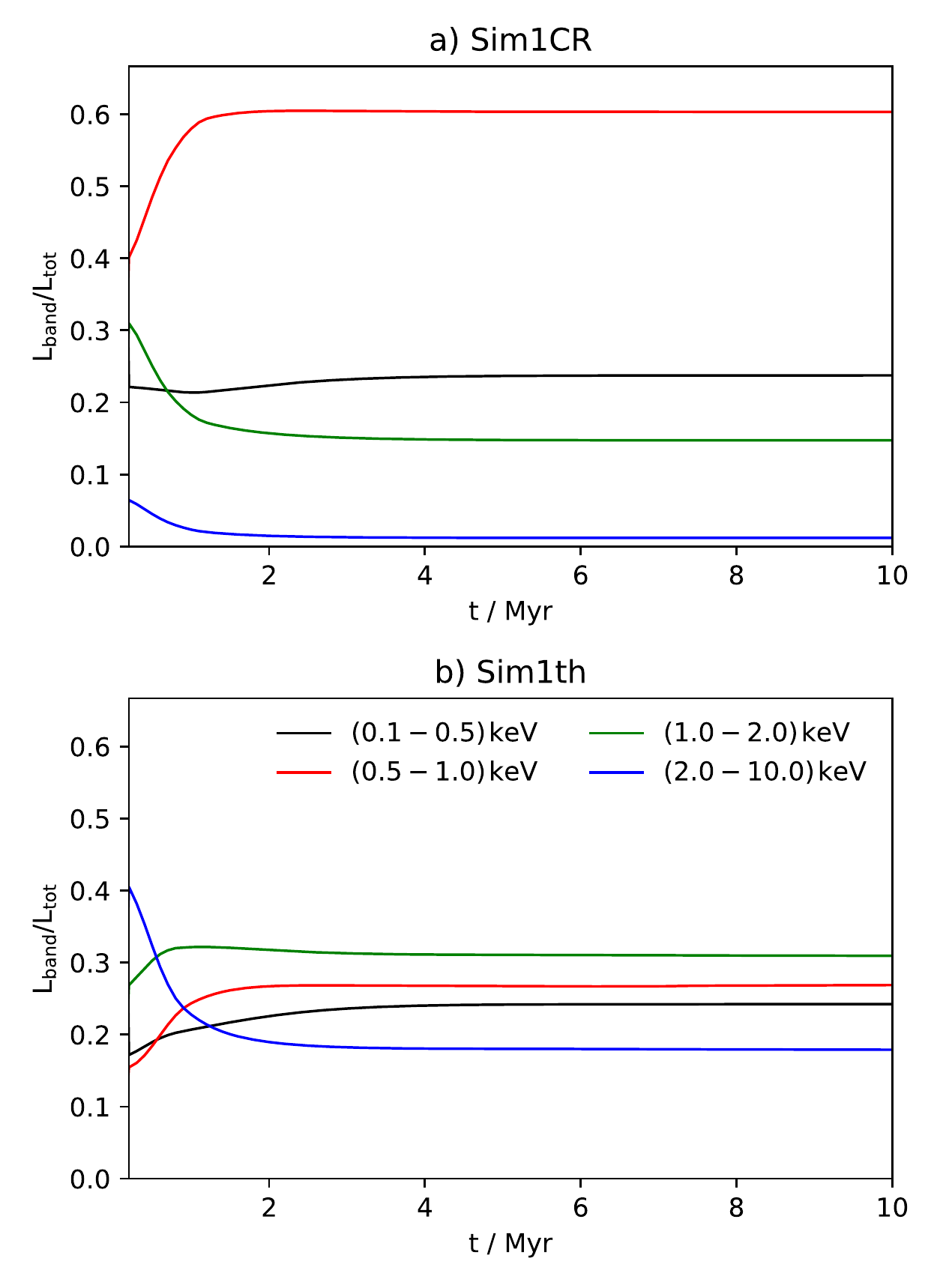} 
\vspace*{-0.25cm}
\caption{Normalised X-ray light curves of the galactic outflows 
  in simulations {\tt Sim1CR} (CR driven) 
  and {\tt Sim1th}(thermal-mechanically driven) 
  in four energy bands: $(0.1-0.5)$~keV, $(0.5-1.0)$~keV, 
  $(1.0-2.0)$~keV and $(2.0-10.0)$~keV. } 
\label{fig:mechanism_broadband}
\end{figure}

\begin{figure*}
\includegraphics[width=0.95\textwidth]{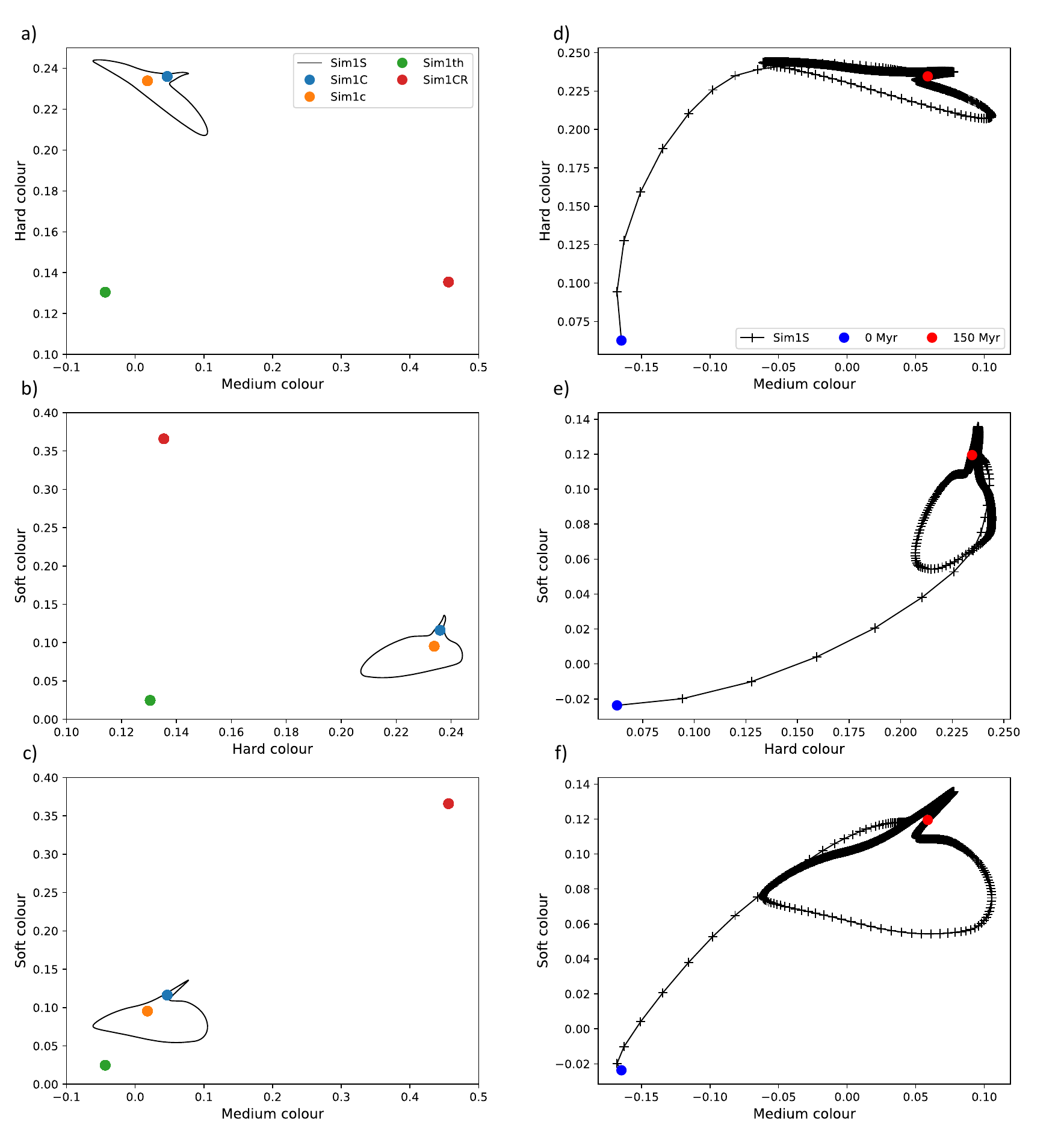} 
\vspace*{-0.25cm}
\caption{3D colour representations, projected onto 3 2D colour axes, of the broadband X-ray emission 
  of the outflows in simulations {\tt Sim1c},  {\tt Sim1C}, 
  {\tt Sim1th} and {\tt Sim1CR}
  after they have reached a stationary state (left panels, a, b and c),  
  and of the outflow in simulation {\tt Sim1S} as it evolves 
  (right panels, d, e and f). The evolution of {\tt Sim1S} is shown between 0 and 150 Myr, with regular 0.1 Myr intervals indicated by the cross marks on the lines in the panels on the right.}  
\label{fig:3d_colour_evolve}
\end{figure*}

\begin{figure*}
\includegraphics[width=0.7\textwidth]{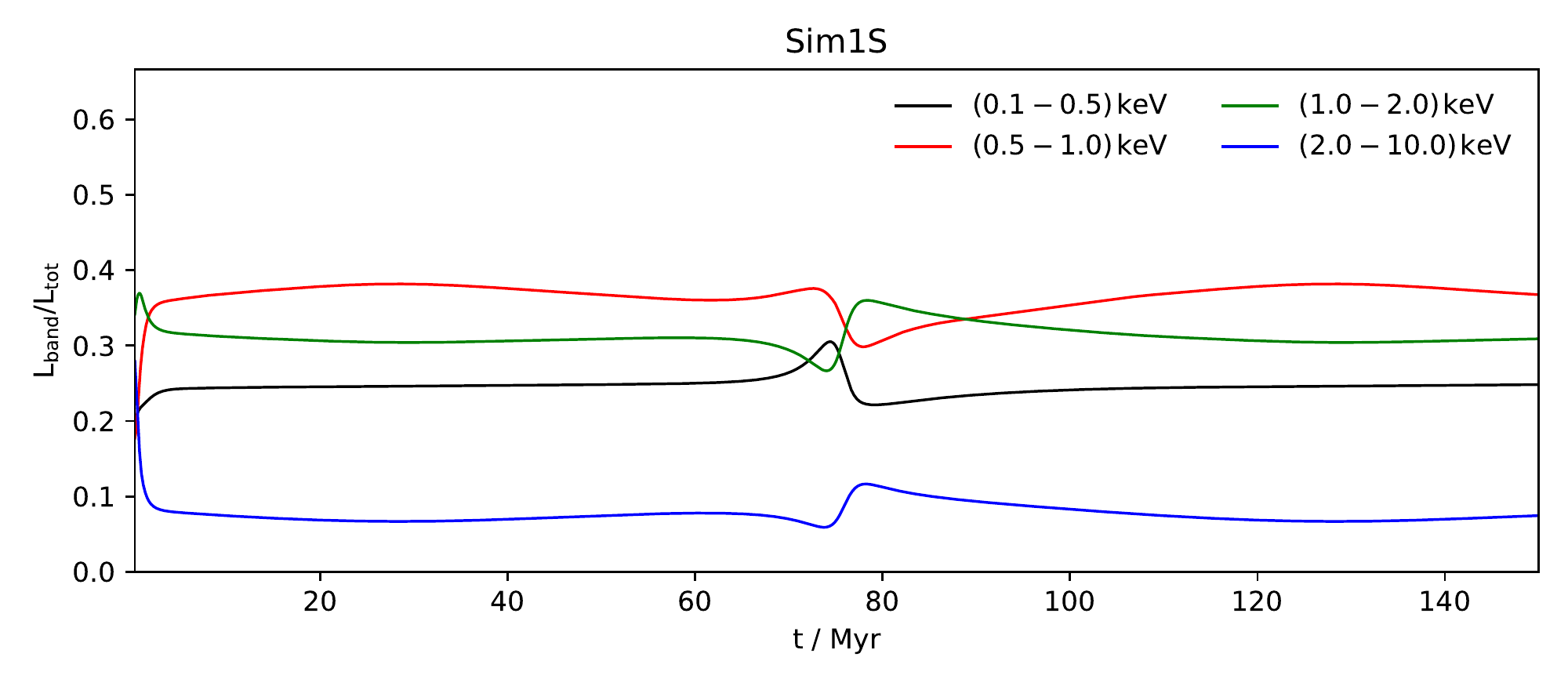} 
\vspace*{-0.25cm}
\caption{Normalised X-ray light curves of the galactic outflow  
  in simulations {\tt Sim1S},  
  in four energy bands: $(0.1-0.5)$~keV, $(0.5-1.0)$~keV, 
  $(1.0-2.0)$~keV and $(2.0-10.0)$~keV.} 
\label{fig:1s_evolve}
\end{figure*}

The large number of photons needed 
  for the construction of a high-resolution X-ray spectrum  
  requires a long observation time 
  on an X-ray telescope  
  with a large effective collecting area.  
Thus, the diagnoses using spectroscopic analyses 
  described in the above subsections,  
  though applicable to nearby galaxies, 
  are not practical in the investigation of outflows 
  from distant galaxies.  
Colour-colour analysis  
  is a more economical alternative,  
  as it requires only broadband information, 
  which is less demanding in terms of the net photon flux.     
We therefore compute the broadband X-ray emission  
  from the simulated galactic outflows, 
  and construct corresponding X-ray light curves. 
Figure~\ref{fig:sim1c_broadband}  
  shows the normalised broadband light curves 
  of the outflows 
  in simulations \verb|Sim1c| (cooling ignored) 
  and \verb|Sim1C| (radiatively cooling included), 
  while Figure~\ref{fig:mechanism_broadband} 
  shows the normalised broadband light curves 
  of the outflows in simulations \verb|Sim1CR| (CR driven) 
  and \verb|Sim1th| (thermal-mechanically driven). 
As shown, 
  the X-ray emission becomes stationary 
  after the outflows reach their stationary state, after about 2~Myr.  
Comparing the stationary state broadband X-ray emission  
  shows obvious contrasts between the outflows in \verb|Sim1CR| (CR driven) 
  and \verb|sim1th| (thermal-mechanically driven). 
The differences between the broadband X-rays 
  for outflows in \verb|sim1c| (cooling ignored) 
  and \verb|sim1C| (cooling included) 
  are not as prominent, but still distinguishable.  
The differences of the broadband X-rays 
  among the simulated outflows 
  are better revealed with a three-colour representation, using: 
  (i) soft colour = 
    band ($0.5 - 1.0$~keV) $-$ band ($0.1 - 0.5$~keV), 
  (ii) medium colour =  
      band ($0.5 - 1.0$~keV) $-$ band ($1.0 - 2.0$~keV), 
      and 
  (iii) hard colour =     
     band ($1.0 - 2.0$~keV) $-$ band ($2.0 - 10.0$~keV).  
The fluxes in the bands above 
  are normalised to the fluxes 
  in the sum of all bands.
In this representation,      
  an outflow from a galaxy at any instant 
  is a phase point in the 3D colour space.  
Figure~\ref{fig:3d_colour_evolve}, left panel, 
  shows the 3D colour-colour-colour plot projected onto 3 2D axes
  for simulated galactic outflows in their stationary states.  
The four outflows are distinguishable by their locations. 

Alternating episodes of star-formation and quenching  
  would introduce cumulative effects 
    on a galactic outflow, 
  and this issue is addressed here 
  using the broadband X-ray emission 
  of the outflow in simulation \verb|Sim1S|  
  (Figure~\ref{fig:1s_evolve})  
  as an illustration.  
Noticeably, 
  the outflow in the second episode of star-formation   
  does not produce the same broadband X-ray light curves seen in the first episode. 
More specifically,  
  the X-ray emission 
  seems not to settle into a stationary state over a timescale of 2~Myr 
  (cf. the light curves 
  in Figures~\ref{fig:sim1c_broadband} and \ref{fig:mechanism_broadband}).  
This can be explained as follows:    
In the first episode of star-formation,    
  the outflow propagates in a pristine ISM, 
  which is uniform and is practically cold 
  while, in the following episode of star-formation,  
  the outflow propagates into the material left-over by the previous outflow activity. 
This left-over material is warm 
  and has non-negligible density and flow velocities. 
Moreover, the material is not uniform  
  and has spatial gradients in its density and pressure. 
As the outflow during the two star-forming episodes  
  does not have identical physical conditions,  
  its X-ray emission, in both time domain and spectral domain,  
  is different. 
This is illustrated in Figure~\ref{fig:3d_colour_evolve}:  
  the track of the outflow in the 3 colour-colour plot projections
  does not trace a closed orbit.  
The failure of the outflow to reach 
  a stationary state in 2~Myr during the second star-forming episode 
  is caused by the introduction of additional time scales and length scales 
  associated with the spatial variations 
  in the warmer, denser material 
  that the outflow is ploughing through in its propagation 
  from the starburst core. 

\section{Discussion} 
\label{sec:discussion}  

\subsection{Outflow simulation configuration and hydrodynamics}
\label{sec:HD_others}

Several recent works have developed sophisticated HD simulations to investigate outflow properties and/or medium substructure in detail, with some invoking 3D realisations, multi-phase media configurations and various treatments of CR physics~\citep[e.g.][]{Tanner2016ApJ, Schneider2018, Schneider2020, Hopkins2021MNRASb, Bustard2021}. While detailed modelling of the internal flow structure and micro-physics fall beyond the scope of this paper, relevant comparisons can still be drawn between the hot wind component considered in those works and the HD properties of the simulations presented in section~\ref{sec:general}.

Our outflow simulations show the initial development of a super-bubble, forward shock and contact discontinuity. These propagate outwards, and the flow settles into a stationary state relatively quickly after the forward shock has passed. The leading edge of the flow traverses the full simulation domain within $\sim$15 Myr, after which the stationary-state configuration is maintained. We typically find outflow terminal velocities of $\sim$800-900$\;$km s$^{-1}$ across all our simulations and, when the stationary state has been attained, density and temperature profiles fall away radially from the starburst core, with central values of around $10^{-25}\;\!{\rm g}\;\!{\rm cm}^{-3}$ and $10^{7}\;{\rm K}$, respectively. While the exact set-up of their simulations is different, this early evolutionary behaviour and the stationary-state HD characteristics are broadly similar to the results shown in~\citet{Schneider2018}, although their adoption of higher mass loading and thermalisation efficiencies yield denser flows with faster terminal velocities, and a slower development. Simulations by \citet{Tanner2016ApJ} and~\cite{Cooper2008ApJ} further showed that the rate at which an outflow develops is slowed by mass-loading, or boosted by energy injection rates via the star-formation rate and/or thermalisation efficiency of energy into the hot wind component.

In their initial models, \citet{Schneider2018} do not include radiative cooling. This matter is considered further in~\citet{Schneider2020} where it was shown that, for mass loading rates similar to those adopted in this work, radiative cooling would not greatly impact the structure of the hot phase of a galactic wind (this is also evident in our results -- see Figure~\ref{fig:HD1D_cooling}, where only moderate differences appear above $\sim$ 4 kpc when radiative cooling is included). Instead, \cite{Melioli2013} demonstrated that this depends strongly on the fraction of mass in a high density wind component compared to the total mass of the galactic wind, and becomes more important in winds where the mass fraction in a cool, dense phase is higher. 

The results shown in~\citealt{Hopkins2021MNRASb} (for simulated galaxy parameters, see~\citealt{Hopkins2021MNRASc}) target lower resolution scales than this work, however the general features seen in their simulations also reflect the trends seen here, with similar flow temperatures and densities at 10 kpc for models where the halo mass is comparable. \citealt{Hopkins2021MNRASb} also showed that, regardless of the exact propagation physics adopted, the presence of CRs in an outflow leads to a cooler and more extended flow. Although this is also consistent with our findings (see Figure~\ref{fig:HD1D_mechanism}) and previous work~\citep[e.g.][]{Girichidis2018}, it has been demonstrated that the exact flow structure and extent is quite sensitive to the detailed CR propagation physics adopted in a simulation~\citep{Wiener2017, Jacob2018, Farber2018, Hopkins2021MNRASb}.

\subsection{Collisional equilibrium and X-ray spectroscopy} 
\label{sec:equilibirum}

Several implicit assumptions have been made 
  in the spectral calculations 
  (\S~\ref{sec:XR_emission}). 
  Among these are that  
  the X-ray emitting gas is in collisional equilibrium,  
  and that the X-ray emission is optically thin from a collisionally-ionised gas. 
The 1D simulated outflows in this study 
  are stratified structures, 
  showing a decrease in their gas density and temperature  
  as the outflows proceed. 
When the variation in the gas density and  temperature 
  is substantial,     
  local collisional equilibrium may not be guaranteed, 
  and this occurs 
  in stratified flows in solar corona, 
  supernova shocks and accretion shocks
  \citep[see e.g.][]{Shapiro1977ApJ,Gorenstein1974ApJG,Wu2001MNRAS}, where thermal equilibrium between ions and electrons     
  and Maxwellian distributions for the energies of the particles  
  may not be maintained. 
The deviation from thermal distribution of the electrons  
  and the deviation from electron-ion collisional equilibrium 
  would affect the bound-bound transitions in the ions 
  and hence their corresponding radiative processes
  \citep{Masai1984Ap&SS,Tatischeff2003EAS,Decaux2003PhRvA,Gu2005ApJ,Cui2019ApJ}. 
The alternation of the spectral line features 
  caused by such deviations 
  would, in turn, affect the reliability 
  of using X-ray spectroscopy 
  to diagnose galactic outflows and their driving mechanisms.  


The collisional time of an ionised gas is  
\begin{align}  
  t_{\rm coll} & \approx 11.4 \ 
  \left( \frac{A^{1/2}\ T^{3/2}}{n_{\rm e}Z^4~(\ln \Lambda)}\right) \  {\rm s} 
\end{align} 
  \citep{Spitzer1956book},  
  where $n_{\rm e}$ is the electron number density (in ${\rm cm}^{-3}$), 
  $T$ is the electron temperature (in ${\rm K}$), 
  $A$ is the mean molecular weight (normalised to hydrogen mass), 
  $Z$ is the ionic charge, 
  and $\ln \Lambda$ is the Coulomb logarithm. 
The outflow at a radial distance $r$, from the starburst core $r_{\rm sb}$,    
  would be able to reach a local collisional equilibrium if 
\begin{align} 
 t_{\rm coll}(r) & < t_{\rm dyn}(r) 
  = \int_{r_{\rm sb}}^{r}\ \frac{{\rm d} \;\!r' }{v(r')} \ , 
\end{align} 
  where $t_{\rm dyn}$ is the dynamical time of the flow, 
  and $v(r')$ is the outflow velocity. 
Specific emissivity of X-ray lines and continuum 
  arising from bound-bound, bound-free and free-free 
  processes generally depend on $\rho^2 T^{3/2}$ 
  \citep[see e.g.][]{Rybicki2004book}, 
  and hence the inner 4~kpc of the outflow would dominate the emission. 
The dynamical time scale of an outflow is $\sim (1-10)~{\rm Myr}$.  
With the densities and temperatures of the simulated outflows 
  (see Figures`\ref{fig:HD1D_Sim2C}, \ref{fig:HD1D_cooling} 
  and \ref{fig:HD1D_mechanism}), 
  the collisional equilibrium condition 
  is satisfied in most of the outflow, 
  except in outermost region 
  where the density drops below $\sim 10^{-27} {\rm g~cm}^{-3}$, 
  corresponding to particle number density 
  of $\sim (10^{-4}- 10^{-3})\;{\rm cm}^{-3}$. 
The gas density in the outflow could be higher in reality,    
  as molecular clumps could fall into the outflows, 
   from which material would be stripped and dissolved into the hot outflow gas, 
  while external gas can also swept into the hot wind fluid.  
While these complications will not substantially distort the emission spectra from the dense, hot inner outflow region, they   
  would add some complexity to the X-ray spectral analysis. 
Molecular clumps and entrained gas 
  are distinguishable 
  by their multi-wavelength spectral signatures 
  and, hence, they also provide 
  additional dimensions in outflow diagnostics. 
  

\subsection{Observational implications}
\label{sec:ob_implication}

X-rays are a useful means 
 of probing the thermal properties of hot gases 
 and, hence, the HD of galactic outflows 
 \citep[see e.g.][]{Strickland2000}. 
Galactic outflows have been imaged in X-rays 
  by {\it Chandra} and {\it XMM-Newton} observations,
  and spatially resolved high-resolution X-ray spectra have been obtained 
  for outflows in several nearby starburst galaxies, 
  e.g. NGC 253 \citep{Mitsuishi2013PASJ} and M82 \citep{Lopez2020}.  
As was shown in \cite{Lopez2020} \citep[see also][]{Strickland2007ApJ}, 
   X-ray brightness is not uniform across the M82 outflows. Instead, it can be seen to decrease over distance from the galactic centre, with spectral variations also being observed. 

Although the synthetic X-ray spectra of the simulated galactic outflows in this work 
  show the same trend as the observations of M82 and other nearby starburst galaxies 
  (regardless of their driving mechanism), 
  there are subtle differences between thermal-mechanically driven and CR driven systems. 
Unambiguously distinguishing these differences 
  will require high quality spectroscopic data 
  and, for galaxies at distances substantially 
  beyond M82 or NGC~253 (beyond $\sim$ 5~Mpc), 
 next generation facilities, such as {\it ATHENA}, will be needed. 
Nonetheless, this work has demonstrated that  
  1D HD simulations and analytic calculations
  are sufficient to provide acceptable model structures, 
  and that post-processed X-ray spectral templates and colour templates 
  can be computed for different outflow models 
  with modest computational effort. 
With the upcoming {\it ATHENA} and other next generation X-ray observatories, 
  we will be able to probe galactic outflows beyond the local Universe 
  with observations, theoretical modelling 
  and even population analyses and stacking analyses 
  (for the investigation of metallicity evolution 
  in the diffuse gas of galactic outflows over redshift). 
 
Outflows from distant galaxies are not always 
 spatially resolved. 
Thus, their X-ray emission is contaminated by stellar sources 
 such as X-ray binaries, 
 and the brightest among these, 
 ULXs (ultra-luminous X-ray sources), 
 could have X-ray luminosities reaching $10^{40}{\rm erg}~{\rm s}^{-1}$  
 \citep[see e.g.][]{Swartz2004ApJS}, 
 i.e. a substantial fraction of the X-ray power of an entire outflow.  
However, X-ray binaries and ULXs
  tend to contaminate spectra in energies above 2~keV 
  \citep[see e.g.][]{Strickland2007ApJ}. 
Their keV X-ray emission also has noticeable Fe~K$\alpha$ lines, 
  with a 6.4~keV neutral component arising from the accretion disk 
  and a 6.97~keV H-like component 
  from the photo-ionised gas in or around the binary. 
The 6.4~keV neutral component 
  and the 6.97~keV H-like component of Fe K$\alpha$ 
  are, however, not present (or extremely weak) in galactic outflow X-ray emission, 
  where the Fe emission is instead dominated by L-shell transitions 
  (see \S~\ref{sec:XR_spectroscopy}).
Note that X-ray binaries and ULXs 
  generally have thermal black-body like emission  
  at around 1~keV, which originates from their accretion disk. 
This thermal black-body like emission at around 1~keV 
  and the properties of Fe K$\alpha$ lines 
  together with survey studies 
  \citep[e.g.][]{Swartz2004ApJS},  
  will provide a means to estimate the X-ray binary and ULX populations 
  \citep[see e.g.][]{Fabbiano2006ARAA,Hua2011NewAR}. 
With the X-ray binary and ULX populations modelled 
  \citep[see][]{Wu2001PASA,Mineo2012MNRAS}, 
  their contribution to the X-rays,  
  including the spectral region of energies below 1~keV, 
  can be estimated, 
  thus allowing the extraction of information 
  about spatially unresolved outflows in distant galaxies 
  using the stacking method, as is also employed in studies 
  of galaxies and AGN in X-rays and other wavebands \citep[e.g.][]{Vito2016MNRAS,Fornasini2018ApJ}.
  
The complex phase structure of outflows, 
   particularly near the starburst region, 
   implies that charge exchange processes 
   would contribute to the X-ray emission lines,  
   together with radiative processes 
   in a collisional ionised plasma. 
Observations \citep[e.g. of the outflows in M82][]{Konami2011,Zhang2014ApJ}
   have shown a substantial contribution of charge exchange 
   in sub-keV emission lines. 
The contamination of emission lines 
  by charge exchange processes appear 
  to be less significant 
  for X-ray lines at higher energies, say above $0.7~{\rm keV}$  
  \citep[see][]{Zhang2014ApJ}. 
  It is unlikely that the Fe L-emission bump, 
  which contributes to most of the X-ray flux 
  at energies around $(0.6-1.0)~{\rm keV}$, 
  would be severely contaminated by the charge exchange emission.   
Moreover, the thermal continuum 
  would dominate the total X-ray flux  
  at energies higher than $\sim 1~{\rm keV}$.
Although charge exchange processes would need to be properly accounted for 
  when line spectroscopy is used for flow diagnosis, 
  analysis using broad-band X-ray colour photometry    
  would be much less affected.
  
For spatially resolved outflows, 
  the Fe K$\alpha$ line in observed spectra  
  could provide a measure of their 
  non-HD aspects. 
For instance, the detection of a neutral 6.4~keV Fe K$\alpha$ line 
  in a spectrum, 
  which requires a substantial amount dense cold material, 
  implies the presence of cold clumps 
  which may indicate condensation due to thermal instabilities 
  and/or strong infall of cold clouds from circum-galactic environments. 
A further confirmation of the cold clumps 
  could be derived from the strengths  
  of charge-exchange lines \cite[see e.g.][]{Zhang2014ApJ, Wu2020}. 
The detection of a H-like 6.97~keV Fe K$\alpha$ line, 
  due to photo-ionisation, 
  implies the presence of a strong hard X-ray radiation field, 
  which might be provided by a weak AGN, 
  or by populations of ULXs in the host galaxy of an outflow.





\subsection{Further remarks} 

There are particular uncertainties involved in modelling the HD of outflows, 
  and therefore caution may be needed 
  when applying X-ray spectral results 
  in practical analyses of outflows from starburst galaxies. 
In particular, three issues  
 -- the treatment of CRs, 
    the amount of mass loading, 
    and the presence of multiple phases in the outflow --
 are worthy of attention and further discussion.  
Here we briefly comment on the effect of each of these. 

How CRs propagate within galaxies 
   remains unsettled.  
 Our limitation in the understanding of 
   the transport of energetic particles 
   in magnetised interstellar and intergalactic media 
   will inevitably introduce uncertainties 
   when we implement CRs  
   as energy-momentum sources  
   contributing to the power of a galactic outflow 
   in a HD formulation.     
This in turn will affect 
   the density and temperature profile 
   of the outflow 
   obtained from HD calculations and simulations 
\citep[see e.g.][]{Ramzan2020ApJ,Hopkins2021MNRASb, Hopkins2021MNRASc}. 
Different CR propagation models can lead to substantial differences 
  in the gas temperatures by as much as an order of magnitude,   
  and the high-density region, 
    where most of the X-ray flux is produced,   
  is often the most affected by uncertainties in CR propagation modelling 
  \citep[see][]{Peters2015,Hopkins2021MNRASb}.   
However, the role of CRs 
  in a HD formulation is to act as agents of energy injection. 
If we can establish the CR calorimetry 
  in the context of energy deposition 
  through an alternative method,    
  it is possible to by-pass theoretical calculations 
  of CR transport, together with their complexities 
  and various uncertainties.   
Baryonic CRs deposit energies into an astrophysical medium 
  via hadronic processes. 
In galactic environments, 
  a major channel is  
  the production of charged and neutral pions 
  \citep[see][]{Owen2018MNRAS}, 
  and the branching ratio is roughly the same 
  for the formation of three pion species \citep[][]{Dermer2009book}. 
The charged pions decay via a weak interaction, generating cascades of 
  less energetic and lighter particles, such as electrons,  
  which are efficient in passing the energy and momentum 
  to the galactic outflow gas. 
Neutral pions decay via an electromagnetic interaction, 
  with a pair of gamma-ray photons being produced 
  \citep{Dermer2009book,Owen2021MNRAS}.  
A galactic outflow will therefore emit gamma-rays    
  if CRs are depositing energy into the outflowing gas 
  \citep{TChan2019MNRAS,Bustard2020ApJ}.  
From the measurement of the luminosities of the gamma-ray glow  
  from the outflows of starburst galaxies,  
  we could empirically derive the source terms associated 
  with CRs in the HD equations, 
  without being hampered by a limited ability  
  to theoretically resolve CR transportation complexities.  
Measuring the gamma-ray glow in external galaxies  
  is one of the science themes 
  of the Cherenkov Telescope Array (CTA)
  \citep{2019sctabook}, 
  and future observations of nearby starburst galaxies,  
  such as the identified CTA targets, NGC 253 and M82, 
  will provide us with an empirical means  
  to model the efficiency of energy and momentum 
  deposition of CRs into galactic outflows. 
Nevertheless, in this study we have shown that 
  for the same energy injection rate, 
  CR-driven outflows 
  are colder than thermally-driven outflows,  
  and this trend is generally insensitive 
  to the CR physics model adopted in the HD formulation 
  \citep[see also][]{Hopkins2021MNRASb}.

While CRs affect the energy and momentum input in an outflow,
  mass loading affects the relative amount of injected energy and momentum 
  shared between the hot ionised gas and the other flow components. 
Although both would alter the density and temperature profiles of outflows 
 \citep{Chevalier1985Nat,Veilleux2005, Martizzi2016}, 
 and hence the spectral properties of their emitted X-rays, 
 the amount of mass loading can be estimated using some X-ray lines,   
 while additional gamma-ray observations  
 would be needed to disentangle the CR involvement in energy deposition. 
More specifically, 
  the variation of mass-loading and thermalisation efficiency 
  would leave observational signatures in X-ray emission line spectra. 
For example, Oxygen can be used as a tracer of mass-loading.   
As shown in the study of \cite{Martin2002ApJ} 
  a galactic wind carries almost all the metals ejected   
  in a starburst driven outflow.  
The Oxygen in the wind is from the stellar ejecta 
  rather than from entrained interstellar gas.  
Oxygen emission lines in X-rays and the derived abundances  
  would therefore offer relatively strong constraints 
  on the mass loading of the hot outflowing component of a wind.  
Mass loading together with thermalization efficiency 
  were determined in previous studies \citep[e,g,][]{Strickland2009ApJ}   
  using the S, Ar and Ca emission lines 
  and the total X-ray flux in the $(2-8)~{\rm keV}$ energy band.    

Galactic outflows are multi-phase in nature   
  \citep[see][]{Aguirre2005,Konami2011,Zhang2014ApJ,Martizzi2016, Li2017, Kim2018, Lopez2020, Schneider2020,Wu2020}. 
The charge-exchange lines observed in the X-ray spectra   
  of outflows from starburst galaxies 
  indicate the presence of neutral clumps \citep{Konami2011,Zhang2014ApJ}. 
Entrainment of these cold clumps would alter the thermal and HD properties 
  of the hot flow component, from where keV X-rays originate.
A cold neutral clump can take away thermal energy  
  from the hot component of a flow 
  when it evaporates. 
At the same time, the hot component would gain mass 
  if the vaporised material from the clumps 
  becomes thermalised and ionised.  
For a clump with a density $\sim 10$ cm$^{-3}$~\citep{Melioli2013}, a minor-axis size of $\sim$ 10 pc and a velocity of $\sim 600~{\rm km~s}^{-1}$ 
  \citep[e.g.][]{Strickland2009ApJ} (comparable to the galactic outflow 
  velocities obtained the HD simulations shown in this work)
   embedded within the hot component of a flow of temperature $\sim 3\times 10^6~{\rm K}$,  
  the evaporation length-scale 
  (and, presumably, the length-scale over which such entrainment would persist) 
  is of order 10s of kpc \citep[see][]{Wu2020}. 
This evaporation length would be longer if the temperature of the outflow gas 
  is cooler. 
The thermalisation of an outflow would therefore not be modified significantly by entrained clumps, 
  even if the velocity of the clumps were lower than assumed here,  
  as the clumps would be advected out 
  to the intergalactic/circumgalactic environment by the outflow before the completion of the thermalisation process.

Although charge-exchange lines are produced 
  at interfaces between cold neutral clumps 
  and the hot flow component, 
  emission lines with energies of 1~keV or higher are not strongly affected, 
  and the keV continuum is practically unaffected.   
The contamination caused by 
  charge-exchange lines from the surfaces of of the cold clumps   
  would not selectively affect the outflows of any particular driving mechanism.  
Thus, the overall trend for X-ray emission 
  across the different driving mechanisms that we have considered  
  is still relatively robust, 
  despite the presence of cold clumps 
  and their associated charge-exchange emission lines in the sub-keV part of the X-ray spectrum.

The most noticeable effect of the presence of the low-temperature gas component 
  is the absorption that it causes at X-ray energies of just below $\sim$ 2~keV. 
Observations can, however, 
  determine the amount of this low-temperature gas  
  by fitting appropriate absorber models, 
  a standard practice in the analysis of X-ray spectral observations. 
The remaining question is now whether or not the X-rays 
  would be severely contaminated by emission from the mass-loading cooler gas 
  with temperatures below $\sim 10^6~{\rm K}$.  
The keV X-ray emission from optically thin 
  collisionally-dominated plasmas satisfying the coronal condition 
  \citep[see e.g.][]{Mewe1981A&AS,Mewe1985A&AS},
  as would be applicable to the hot gas in galactic outflows,   
  is determined by the density and temperature of the emitting gas. 
The specific flux of the emission from a plasma in a coronal condition is given by 
\begin{align}  
  F_\nu & =  \rho^2 V \ f_\nu(Z_{\rm i},T,n_{\rm l},n_{\rm u}) \ ,  
\end{align}
 where $\rho$ is the mass density, $V$ is the volume of the emission region, 
 and $f_\nu(Z_{\rm i},T,n_{\rm l},n_{\rm u})$ is a function of thermal temperature ($T$),  
   the charges of the ion species ($Z_{\rm i}$), and 
   the populations of electrons in the upper level and lower level 
   of the transition ($n_{\rm u}$ and $n_{\rm l}$ respectively).  
Generally, 
 $f_\nu \propto (k_{\rm B}T)^{-[\alpha+(3/2)]}  \exp(-h\nu/k_{\rm B}T)$, 
 where $h$ is the Planck constant, $k_{\rm B}$ is the Boltzmann constant, 
 and $\alpha = -1.68$, $-2$ and $-1$ 
 for bound-bound, bound-free and free-free processes respectively  
 \citep{Jefferies1968book,Rybicki2004book}.  
Because of the $\exp(-h \nu/k_{\rm B} T)$ drop-off, 
  only gas with temperatures of $T \sim 10^7{\rm K}$ or above
  will make a substantial contribution 
  to the keV part of the X-ray spectrum, 
  while gas with lower temperatures will not make a significant contribution 
  in collision-induced radiative processes.  
Similarly, gas with temperatures lower than $\sim 10^6{\rm K}$ 
  will not make a significant contribution to the $(0.1-1)$~keV part of the spectrum. 
This result is also reflected in the flux ratio plot in Figure~\ref{fig:dLxdT_regions}, 
  which shows that the contribution of gas with a temperature below $10^6~{\rm K}$  
  to the keV X-ray is practically zero.


\section{Conclusions}
\label{sec:conclusion}
 
In this work, we investigate the HD and X-ray emission properties of galactic outflows, driven by thermal-mechanical energy, CRs and their mix. We use simulations to investigate their HD profiles and properties as they evolve from their initial condition to a stationary state, and find their stationary-state configurations to be consistent with those previously determined analytically, in~\citetalias{Yu2020}. We compute synthetic X-ray spectra from the HD properties of the flows to assess their observable properties, and demonstrate how broadband X-ray analyses can be used to discriminate between outflows according to their principal driving mechanism at different stages of their evolution. We find that the keV X-ray emission of the simulated outflows are dominated by the inner $\sim$4 kpc of the flow, where the collisional equilibrium timescale would be substantially shorter than the dynamical time of the outflow. The continuum emission is
mainly determined by the density and temperature, with minimal dependence on other factors, while 
spectral features (in particular the Fe K$\alpha$ line and L emission complex, the He-like K$\alpha$ emission from O, and the N K$\alpha$ emission) are more strongly impacted by the flow metallicity, as well as being sensitive to temperature.

Outflows from starburst galaxies in the nearby Universe (such at M82) have been studied extensively, and their emission properties can easily be resolved in observations over a short integration time. However, outflows in more distant galaxies are not as easy to access. 
Galactic outflows are a consequence of star-forming activity, and their properties and characteristics are inextricably linked to the evolution of their host galaxy~\citep{Mannucci2010}. As such, the properties of outflows would show evolutionary trends over redshift~\citep{Sugahara2019ApJ}, reflecting those seen in their parent galaxy populations~\citep{Madau1996, Dickinson2003}. We have demonstrated that this variation between flows and their evolution can be captured observationally by carefully selected broad X-ray bands, which would retain the crucial information necessary to probe their internal HD and composition (including metallicity), while avoiding the need for the construction of a high-resolution spectrum with long observational integration times and large effective collecting areas. 
Such an approach therefore opens up the possibility of studying the internal physical properties of outflows into the distant Universe, allowing the importance of different galactic outflow driving mechanisms and internal properties to be studied 
over cosmic time, reaching systems in the earlier Universe, e.g. during the cosmic noon, when conditions and physical processes in and around galaxies would be very different to those seen today.

\section*{Acknowledgements} 

BPBY's visit to National Tsing Hua University (NTHU)  
  was supported by an NTHU international exchange scholarship, 
  and by the Ministry of Science and Technology of Taiwan (ROC) 
  through grant 107-2112-M-007-032-MY3. 
ERO is supported by the Ministry of Education of   
  ROC (Taiwan) at the NTHU Center for Informatics and Computation in Astronomy (CICA). 
This research made use of the high-performance computing facilities at CICA, 
  operated by the NTHU Institute of Astronomy 
  and funded by the Ministry of Education and 
  the Ministry of Science and Technology 
  of Taiwan (ROC).  
This research has also made use of NASA's Astrophysics Data Systems. 
The software \verb|FLASH| used in this work was in part 
  developed by the DOE NNSA-ASC OASCR Flash Center at the University of Chicago. We thank the anonymous referee for their comments, which helped to improve the manuscript. 

\section*{Data Availability}
The data underlying this article will be shared on reasonable request to the corresponding author.


\bibliographystyle{mnras}
\interlinepenalty=10000
\bibliography{Brian2020MNRAS}



\appendix

\section{Comparison between numerical simulations and analytic models}
\label{sec:comparison_analytic}

\begin{figure*}
\includegraphics[width=0.85\textwidth]{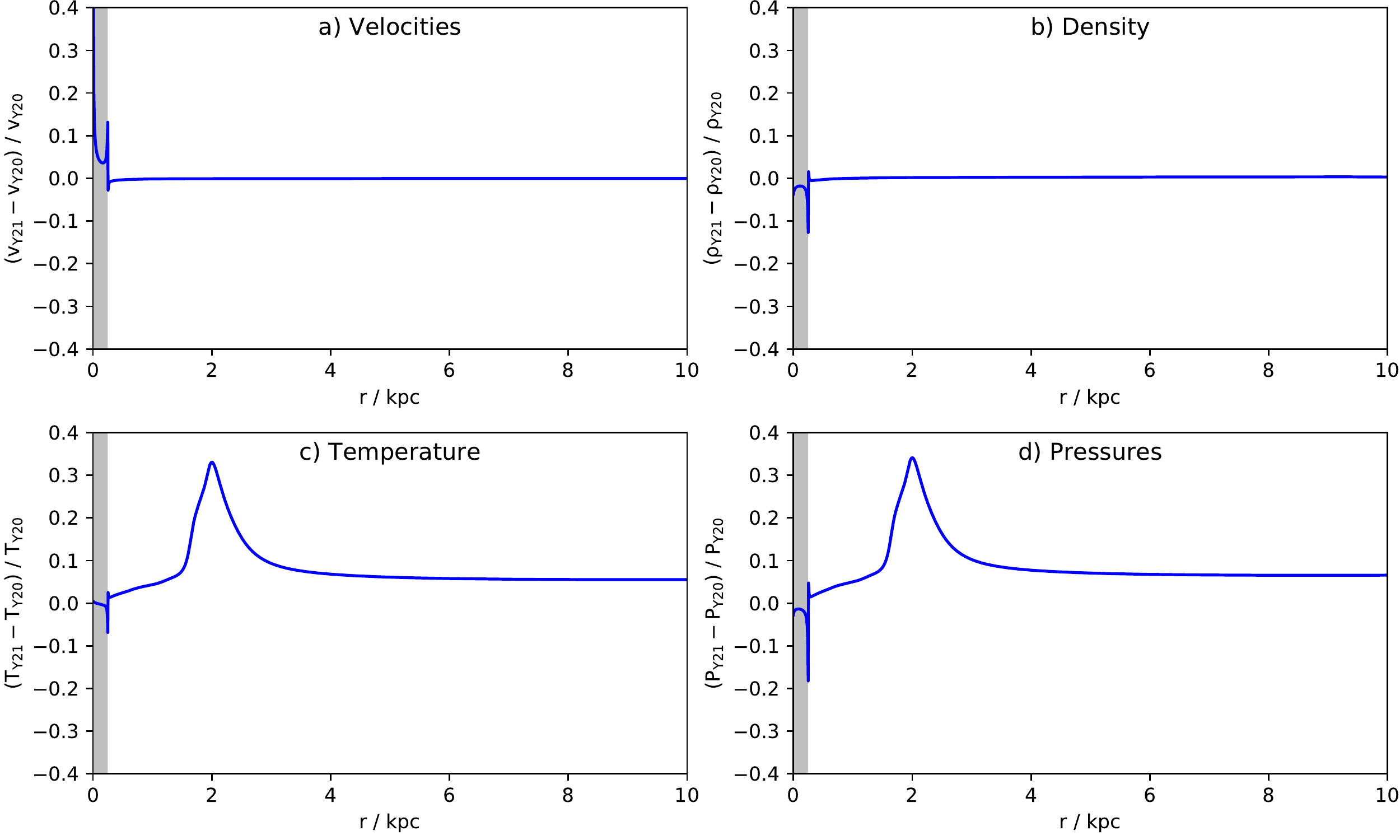}
\caption{Fractional discrepancies between the stationary-state HD profiles from numerical simulations in this paper and analytical calculation in \citetalias{Yu2020}, where the same set of parameters is adopted. Outflow velocities, gas density, gas temperature, and gas pressure are plotted in panel \textit{a}, \textit{b}, \textit{c}, and \textit{d} respectively. The starburst region is denoted by the shaded area at $r<250$ pc, and the comparison is shown up to 10 kpc to match the size of the simulation domain.}
\label{fig:HD1D_appendix}
\end{figure*}

The asymptotic stationary-state HD profiles computed in our numerical simulations are compared with the analytical stationary-state outflow solution of \citetalias{Yu2020} in Figure \ref{fig:HD1D_appendix}.
The fractional residuals in the velocity, density, temperature and pressure profiles show that there is excellent agreement between the stationary-state results of this work and those of \citetalias{Yu2020}, if the same set of system parameters is adopted.
We note that the minor differences seen in the temperature and pressure profiles at around $\sim$2 kpc arise due to radiative cooling, which is strongest at these locations.


\bsp	
\label{lastpage}
\end{document}